\documentclass[twoside,twocolumn,9pt,hidelinks]{article}
\usepackage{extsizes}
\usepackage[super,sort&compress,comma]{natbib} 
\usepackage[version=3]{mhchem}
\usepackage[left=1.5cm, right=1.5cm, top=1.785cm, bottom=2.0cm]{geometry}
\usepackage{balance}
\usepackage{mathptmx}
\usepackage{sectsty}
\usepackage{graphicx} 
\usepackage{lastpage}
\usepackage[format=plain,justification=justified,singlelinecheck=false,font={stretch=1.125,small,sf},labelfont=bf,labelsep=space]{caption}
\usepackage{float}
\usepackage{fancyhdr}
\usepackage{fnpos}
\usepackage[english]{babel}
\addto{\captionsenglish}{%
  
}
\usepackage{array}
\usepackage{droidsans}
\usepackage{charter}
\usepackage[T1]{fontenc}
\usepackage[usenames,dvipsnames]{xcolor}
\usepackage{setspace}
\usepackage[compact]{titlesec}
\usepackage{hyperref}
\usepackage{siunitx}
\usepackage{layouts}
\usepackage{amssymb}
\usepackage{bm}
\usepackage{color,soul}
\usepackage{epstopdf}%This line makes .eps figures into .pdf - please comment out if not required.

\definecolor{cream}{RGB}{222,217,201}

\begin{document}

\pagestyle{fancy}
\thispagestyle{plain}
\fancypagestyle{plain}{
%%%HEADER%%%
\renewcommand{\headrulewidth}{0pt}
}
%%%END OF HEADER%%%

%%%PAGE SETUP - Please do not change any commands within this section%%%
\makeFNbottom
\makeatletter
\renewcommand\LARGE{\@setfontsize\LARGE{15pt}{17}}
\renewcommand\Large{\@setfontsize\Large{12pt}{14}}
\renewcommand\large{\@setfontsize\large{10pt}{12}}
\renewcommand\footnotesize{\@setfontsize\footnotesize{7pt}{10}}
\makeatother

\renewcommand{\thefootnote}{\fnsymbol{footnote}}
\renewcommand\footnoterule{\vspace*{1pt}% 
\color{cream}\hrule width 3.5in height 0.4pt \color{black}\vspace*{5pt}} 
\setcounter{secnumdepth}{5}

\makeatletter 
\renewcommand\@biblabel[1]{#1}            
\renewcommand\@makefntext[1]% 
{\noindent\makebox[0pt][r]{\@thefnmark\,}#1}
\makeatother 
\renewcommand{\figurename}{\small{Fig.}~}
\sectionfont{\sffamily\Large}
\subsectionfont{\normalsize}
\subsubsectionfont{\bf}
\setstretch{1.125} %In particular, please do not alter this line.
\setlength{\skip\footins}{0.8cm}
\setlength{\footnotesep}{0.25cm}
\setlength{\jot}{10pt}
\titlespacing*{\section}{0pt}{4pt}{4pt}
\titlespacing*{\subsection}{0pt}{15pt}{1pt}
%%%END OF PAGE SETUP%%%

%%%FOOTER%%%
\fancyfoot{}
%\fancyfoot[LO,RE]{\vspace{-7.1pt}\includegraphics[height=9pt]{head_foot/LF}}
%\fancyfoot[CO]{\vspace{-7.1pt}\hspace{13.2cm}\includegraphics{head_foot/RF}}
%\fancyfoot[CE]{\vspace{-7.2pt}\hspace{-14.2cm}\includegraphics{head_foot/RF}}
\fancyfoot[RO]{\footnotesize{\sffamily{\thepage /\pageref{LastPage}}}}
\fancyfoot[LE]{\footnotesize{\sffamily{\thepage /\pageref{LastPage}}}}
\fancyhead{}
\renewcommand{\headrulewidth}{0pt} 
\renewcommand{\footrulewidth}{0pt}
\setlength{\arrayrulewidth}{1pt}
\setlength{\columnsep}{6.5mm}
\setlength\bibsep{1pt}
%%%END OF FOOTER%%%

%%%FIGURE SETUP - please do not change any commands within this section%%%
\makeatletter 
\newlength{\figrulesep} 
\setlength{\figrulesep}{0.5\textfloatsep} 

\newcommand{\topfigrule}{\vspace*{-1pt}% 
\noindent{\color{cream}\rule[-\figrulesep]{\columnwidth}{1.5pt}} }

\newcommand{\botfigrule}{\vspace*{-2pt}% 
\noindent{\color{cream}\rule[\figrulesep]{\columnwidth}{1.5pt}} }

\newcommand{\dblfigrule}{\vspace*{-1pt}% 
\noindent{\color{cream}\rule[-\figrulesep]{\textwidth}{1.5pt}} }

\makeatother
%%%END OF FIGURE SETUP%%%

%%%TITLE, AUTHORS AND ABSTRACT%%%
\twocolumn[
  \begin{@twocolumnfalse}
%{\includegraphics[height=30pt]{head_foot/journal_name}\hfill\raisebox{0pt}[0pt][0pt]{\includegraphics[height=55pt]{head_foot/RSC_LOGO_CMYK}}\\[1ex]
%\includegraphics[width=18.5cm]{head_foot/header_bar}}\par
\vspace{1em}
\sffamily
%\begin{tabular}{m{4.5cm} p{13.5cm} }

%\includegraphics{head_foot/DOI} & 
\noindent\LARGE{%\textbf{Micropump and micromixer capable of cell focusing and trapping$^\dag$}
\textbf{Sharp-edge-based acoustofluidic chip for programmable pumping, mixing, cell focusing and trapping$^\dag$}
}
\vspace{0.1cm}% & \vspace{0.3cm} \\

 \noindent\large{Alen Pavlic,\textit{$^{a\ddag}$} Cooper Lars Harshbarger,\textit{$^{a,b,c}$}, Luca Rosenthaler,\textit{$^{a}$} Jess Gerrit Snedeker,\textit{$^{b,c}$} and J{\"u}rg Dual\textit{$^{a}$}} %Author names go here instead of "Full name", etc.

\vspace{0.6cm}

\noindent\normalsize{Precise manipulation of fluids and objects on the micro scale is seldom a simple task, but nevertheless crucial for many applications in life sciences and chemical engineering.
%We present a micropump fabricated in a silicon-glass chip that is driven by a pair of acoustically excited sharp edges that produce a strong mixing flow in their vicinity.
We present a microfluidic chip fabricated in silicon-glass, featuring one or several pairs of acoustically excited sharp edges at side channels that drive a pumping flow throughout the chip and produce a strong mixing flow in their vicinity.
The chip is simultaneously capable of focusing cells and microparticles that are suspended in the flow. The multifunctional micropump provides a continuous flow across a wide range of excitation frequencies ($\SI{80}{\kilo\hertz}-\SI{2}{\mega\hertz}$), with  flow rates ranging from $\SI{}{\nano\liter\per\minute}$ to $\SI{}{\micro\liter\per\minute}$, depending on the excitation parameters. In the low-voltage regime, the flow rate depends quadratically on the voltage applied to the piezoelectric transducer, making the pump programmable. The behaviour in the system is elucidated with finite element method simulations, which are in good agreement with experimentally observed behaviour. The acoustic radiation force arising due to a fluidic channel resonance is responsible for the focusing of cells and microparticles, while the streaming produced by the pair of sharp edges generates the pumping and the mixing flow. If cell focusing is detrimental for a certain application, it can also be avoided by exciting the system away from the resonance frequency of the fluidic channel. The device, with its unique bundle of functionalities, displays great potential for various bio-chemical applications.%The pump is designed in a form of a silicon-glass chip, offering a relatively cheap batch production.
}

%\end{tabular}

 \end{@twocolumnfalse} \vspace{1.0cm}

  ]
%%%END OF TITLE, AUTHORS AND ABSTRACT%%%

%%%FONT SETUP - please do not change any commands within this section
\renewcommand*\rmdefault{bch}\normalfont\upshape
\rmfamily
\section*{}
\vspace{-1cm}

%%%FOOTNOTES%%%

\footnotetext{\textit{$^{a}$~Institute for Mechanical Systems, Swiss Federal Institute of Technology Zurich, Zurich, Switzerland.}}
\footnotetext{\textit{$^{b}$~Department of Orthopedics, Balgrist University Hospital Zurich, University of Zurich, Zurich, Switzerland.}}
\footnotetext{\textit{$^{c}$~Institute for Biomechanics, Swiss Federal Institute of Technology Zurich, Zurich, Switzerland.}}

%Please use \dag to cite the ESI in the main text of the article.
%If you article does not have ESI please remove the the \dag symbol from the title and the footnotetext below.
\footnotetext{\dag~Supplemental material available: Experimental videos; details on the numerical model; admittance analysis; details on measuring the acoustic pressure amplitude.}
%additional addresses can be cited as above using the lower-case letters, c, d, e... If all authors are from the same address, no letter is required
\footnotetext{\ddag~E-mail: apavlic@ethz.ch}

%%%END OF FOOTNOTES%%%

%%%MAIN TEXT%%%%
\section*{Introduction}

Biomedical and chemical processes often require small amounts of bio-chemical substances, including cell and microparticle suspensions, to be moved around at the micro scale with great precision.\cite{wang2018micropumps,byun2014pumps} There are various mechanisms that have been developed towards this objective, sometimes employing physical principles of piezoelectricity in combination with a diaphragm,\cite{dereshgi2021piezoelectric} electrostatic\cite{kim2014integrated} and electromagnetic\cite{rusli2018electromagnetic} forces, thermal actuation,\cite{hamid2017stack} bubble expansion and collapse,\cite{oskooei2015bubble} electrowetting,\cite{khoshmanesh2017liquid} and acoustics,\cite{huang2014reliable,wang2010valveless,ryu2010micropumping,nabavi2008analysis,ozcelik2021practical,lin2019acoustofluidic,jiang2021rapid} among others.

\begin{figure}[h]
 \centering
 \includegraphics{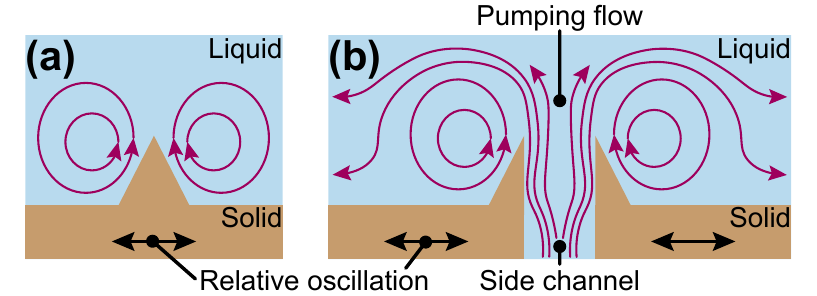}
 \caption{Visualization of (a) the typical acoustic streaming flow around a single sharp edge that oscillates relative to the liquid, and of (b) the pumping mechanism driven by a pair of sharp edges at a side channel.}
 \label{fig:physics}
\end{figure}

%Background - streaming
One particular way of enforcing the fluid motion in microfluidic systems is through a phenomenon called acoustic streaming.\cite{lighthill1978acoustic} This streaming is a steady fluid motion due to the attenuation of acoustic oscillations near a boundary\cite{rayleigh1884circulation,schlichting1932berechnung} or in the bulk of the fluid,\cite{eckart1948vortices} but can also stem from a dynamic geometric nonlinearity.\cite{zhang2021powerful} A special form of the boundary-driven streaming is the streaming near sharp edges, which can be traced back to the acoustic needle experiments of \citet{hughes1962cell} in 1962, for the disruption of cells. Since then, acoustically excited sharp edges have proven to be a promising technology for streaming-based mixing\cite{zhang2019acoustic,zhang2020acoustic,zhang2020unveiling,huang2013acoustofluidic,nama2016investigation,bachman2020acoustofluidic,doinikov2020acoustic} and pumping\cite{huang2014reliable} of fluids, and even for acoustic-radiation-force-based trapping of particles and cells.\cite{leibacher2015acoustophoretic,doinikov2020arf} A series of recent studies revealed a general streaming flow pattern around a single sharp edge that features an outflowing vortex at the sharp tip, as shown in Fig. \ref{fig:physics}(a).\cite{ovchinnikov2014acoustic,doinikov2020acoustic,zhang2020unveiling,zhang2019acoustic,zhang2020acoustic}

%Background - ARF
In addition, the acoustic radiation force (ARF) is used for particle and cell trapping and focusing,\cite{evander2012acoustofluidics} as well as for the separation of particles, cells, and droplets.\cite{wu2019acoustofluidic,leibacher2015microfluidic} A suitable ARF for such applications can be achieved in different systems, one of which is the bulk acoustic wave (BAW) device.\cite{bruus2011forthcoming} When a BAW device is excited, the acoustic waves propagate throughout the whole device, forming standing waves when the frequency of excitation corresponds to a certain resonance mode of the fluidic channel or the whole device.\cite{ozer2021extended,moiseyenko2019whole}

Devices featuring such acoustic phenomena are recognized, for example, as a superior alternative to passive microreactors for chemical engineering.\cite{CHEN2021133258} However, one of the main challenges for the improvement of the technology was defined as constructing the acoustic microreactors from materials other than polydimethylsiloxane (PDMS) and lithium niobate, which would reduce the cost and production complexity while increasing the resistance to extreme reaction conditions.

%What do we do?
Here, we present a silicon-glass chip which is driven by a pair of acoustically excited sharp edges and is simultaneously capable of fluid pumping and mixing, and of focusing and trapping of cells and microparticles in a suspension. The robust device provides a continuous flow across a wide range of excitation frequencies ($\SI{80}{\kilo\hertz} - \SI{2}{\mega\hertz}$), and programmable flow rates in the \SI{}{\nano\liter\per\minute} to \SI{}{\micro\liter\per\minute} range. The novel pumping mechanism featured in our device is based on a pair of sharp edges located at a side channel, pumping the liquid out of the side channel, as depicted in Fig. \ref{fig:physics}(b). Furthermore, in contrast to a low-frequency sharp-edge-based pump made of polydimethylsiloxane (PDMS) by \citet{huang2014reliable}, where the solid edges oscillate and drive the flow, our device operates at higher frequencies that support resonances in the solid-liquid system. The resonances can boost the relative oscillatory velocity between sharp edges and the liquid, which is known to drive the streaming,\cite{doinikov2020acoustic} and consequently the pumping. The behaviour in the experimental system is explained through finite element method simulations. It is shown that the streaming around sharp edges drives the pumping and the mixing, while using a specific frequency can excite a standing wave in a part of the microfluidic device, leading to a simultaneous ARF-based focusing or trapping of cells and microparticles. Exciting the system at an off-resonance frequency of the fluidic channel, allows for the suppression of the focusing, thus allowing for the particles to remain dispersed in the pumping flow if desired.

The silicon-glass composition makes our device chemically resistant, solving one of the challenges of acoustic microreactors for chemical engineering.\cite{CHEN2021133258} Furthermore, our device operates in a relatively broad frequency range, from \SI{80}{\kilo\hertz} to \SI{2}{\mega\hertz}, enabling the control of acoustic cavitation, which is interesting for biomedical applications such as cell lysis.\cite{wiklund2012acoustofluidics} The additional functionalities of our device \textemdash mixing, focusing, and trapping \textemdash make it attractive for various applications, such as boosting the sensitivity of bead-based immunoassays or bioassays,\cite{wiklund2013acoustofluidics} trapping cells for medium exchange,\cite{gerlt2021acoustofluidic} or improving cell isolation.\cite{jiang2021rapid}

%\textit{Interesting/main results?} Main results include:
%\begin{itemize}
%    \item We can pump and focus simultaneously (show that we can also just pump, with $\SI{80}{\kilo\hertz}$ excitation). We can also just pump, without focusing.
%    \item We show a robust micropump that works across a wide range of frequencies, which we demonstrate experimentally and numerically. At specific frequencies, standing waves can be excited in specific parts of the microfluidic channel, 
%    leading to simultaneous focusing or trapping of particles/cells. We also show that there are different ways to reach desired pumping performance/flow rate; many geometrical parameters can be adjusted as well as simply the applied voltage.
%    \item Here, focus just on showing the pumping+focusing and showing the PT simulation for basic parameter exploration.
%\end{itemize}

%Why is it important?

\section*{Theoretical background}

%%% THEORETICAL BACKGROUND %%%

%Here, all the governing equations and other equations related to the sharp edges, ARF, AS, and microfluidic flow are given:
%\begin{itemize}
%    \item Basic set of equations and boundary conditions
%    \item Perturbation theory and resulting equations
%    \item Gor'kov potential, acoustic energy density and ARF
%    \item Any scaling laws that were found regarding sharp edges and other qualitative things about the sharp edges (maybe a detailed overview)
%\end{itemize}

The motion of a viscous fluid is governed by the compressible Navier-Stokes equations
\begin{equation}
    \rho \left[ \frac{\partial \boldsymbol{v}}{\partial t} + ( \boldsymbol{v} 
    \boldsymbol{\cdot} \boldsymbol{\nabla} ) \boldsymbol{v} \right] = - \boldsymbol{\nabla} p + \eta \nabla^2 \boldsymbol{v} + \left( \eta_{\mathrm{B}} + \frac{\eta}{3} \right) \boldsymbol{\nabla} \left( \boldsymbol{\nabla} \boldsymbol{\cdot} \boldsymbol{v} \right) \label{al:NS} ,
\end{equation}
and the continuity equation
\begin{equation}
    \frac{\partial \rho}{\partial t} = -\boldsymbol{\nabla} \boldsymbol{\cdot} \left( \rho \boldsymbol{v} \right) \label{al:cont} ,
\end{equation}
with the dynamic viscosity $\eta$ and the bulk viscosity $\eta_{\mathrm{B}}$. The density $\rho$ is assumed to only be a function of the pressure $p$, namely
\begin{equation}
    \rho = \rho(p) .
\end{equation}

The equations are linearized using the regular perturbation approach.\cite{bruus2012perturbation} Accordingly, the physical fields are expanded in a series, $\square = \square_0 + \square_1 + \square_2 + \dots$, with $\square$ representing the field and the subscript representing the respective order. %We assume the amplitude of the first-order velocity $\vect{v}_1$ to be small with respect to the speed of sound $c_{\mathrm{f}}$ (small Mach number assumption).
The first- and second-order problems, for computing the acoustic fields and streaming fields, respectively, are described in detail in the supplemental material.$^\dag$ The second-order problem becomes relevant only when time scales longer than the period of oscillation $T=1/f$ are considered. Therefore, the time average $\langle \square_2 \rangle = 1 / T \int_T \square_2 \, \mathrm{d}t $ is applied to the second order variables and terms.

\subsection*{Acoustic radiation force (ARF)}
When an acoustic wave inside an acoustofluidic device scatters at an object such as a microparticle or a cell, it gives rise to the acoustic radiation force (ARF). The ARF results from the interactions between the scattered and the background acoustic waves. When an inviscid fluid and a spherical object can be assumed, which is often the case, the ARF can be approximated as the negative gradient of the Gor'kov potential,\cite{gor1962forces} namely
\begin{equation}
  \boldsymbol{F}_\mathrm{rad} = -\boldsymbol{\nabla} U,
  \label{eq:Frad}
\end{equation}
where the Gor'kov potential can be written as
\begin{equation}
  U=\frac{4}{3} \pi a^3 \left( \frac{1}{2} \frac{f_1}{ c_0^2 \rho_0} \langle p_1^{\mathrm{bg}} p_1^{\mathrm{bg}} \rangle - \frac{3}{4} \rho_0 f_2 \langle \boldsymbol{v}_1^{\mathrm{bg}} \boldsymbol{\cdot} \boldsymbol{v}_1^{\mathrm{bg}} \rangle \right),
  \label{eq:Gor'kov}
\end{equation}
with the object's radius being $a$, the speed of sound in the fluid $c_0$, the equilibrium density $\rho_0$, the background acoustic pressure $p_1^{\mathrm{bg}}$, the background acoustic velocity $\boldsymbol{v}_1^{\mathrm{bg}}$, and the monopole and the dipole scattering coefficients $f_1$ and $f_2$, respectively. This formulation is applicable under the assumption that the object is small compared to the acoustic wavelength and far away from the walls. However, a recent study\cite{baasch2020acoustic} showed that the Gor'kov potential is in most cases valid even in close proximity to a wall.

When considering a one-dimensional plane standing wave along a $z$-axis, the ARF from eq. \ref{eq:Frad} further simplifies to
\begin{equation}
  \boldsymbol{F}_\mathrm{rad}^{\mathrm{1D}} = 4 \pi a^3 \Phi k E_{\mathrm{ac}} \sin \left(2 k z \right) \boldsymbol{e}_z,
  \label{eq:Frad1D}
\end{equation}
with the acoustic energy density $E_{\mathrm{ac}} = \frac{p_{\mathrm{a}}^2}{4 \rho_0 c_0^2}$, the pressure amplitude $p_{\mathrm{a}}$, the ideal wave number $k = \frac{\omega}{c_0}$, and the unit vector $\boldsymbol{e}_z$ oriented along the $z$-axis. Equation (\ref{eq:Frad1D}) is attributed to \citet{yosioka1955acoustic}. The acoustic properties of the object relative to the fluid are condensed into the acoustic contrast factor
\begin{equation}
    \Phi = \frac{1}{3}f_1+\frac{1}{2}f_2 = \frac{1}{3}\left[\frac{5\Tilde{\rho}-2}{2\Tilde{\rho}+1} - \Tilde{\kappa}\right],
     \label{eq:Phi}
\end{equation}
in which the relative compressibility $\Tilde{\kappa} = \frac{\kappa_{\mathrm{obj}}}{\kappa_0} $ and density $\Tilde{\rho} = \frac{\rho_{\mathrm{obj}}}{\rho_0}$ reflect the ratios between the properties of the object $\square_{\mathrm{obj}}$ and the fluid $\square_0$.

\begin{figure}[h]
 \centering
 \includegraphics{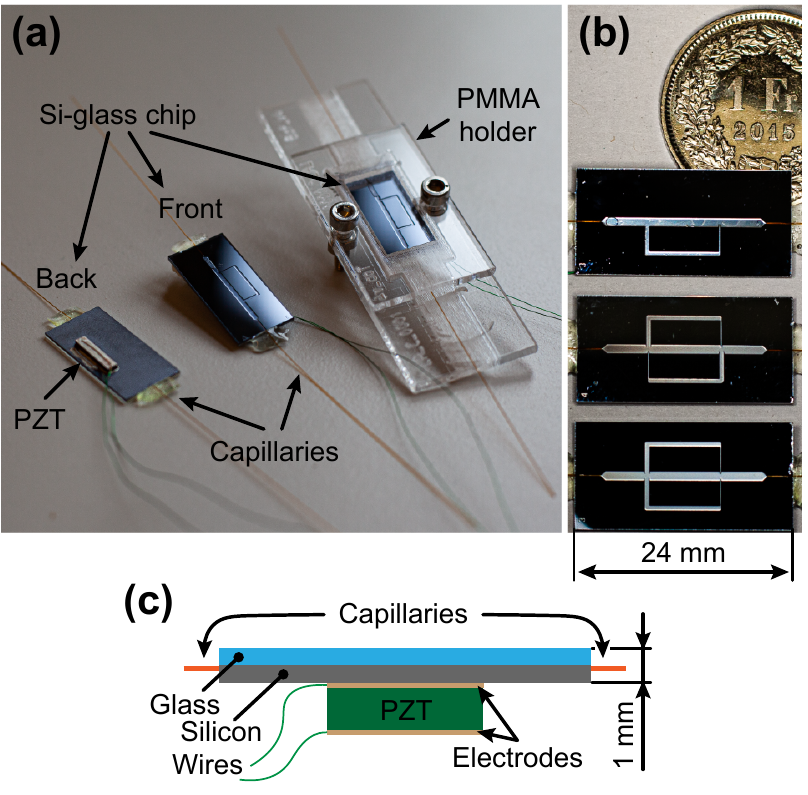}
 \caption{(a) Experimental device, featuring a silicon-glass chip, piezoelectric transducer (PZT), holder from acrylic glass (PMMA), and connections made out of fused silica capillaries. (b) Three different device designs that are featured in the manuscript next to a one Swiss franc coin for scale. (c) The schematic cross-section of a device, explaining individual layers. The size of an individual silicon-glass chip without supporting elements is $1\times12\times\SI{24}{\milli\meter}$.}
 \label{fig:devices}
\end{figure}

\begin{figure*}
 \centering
 \includegraphics{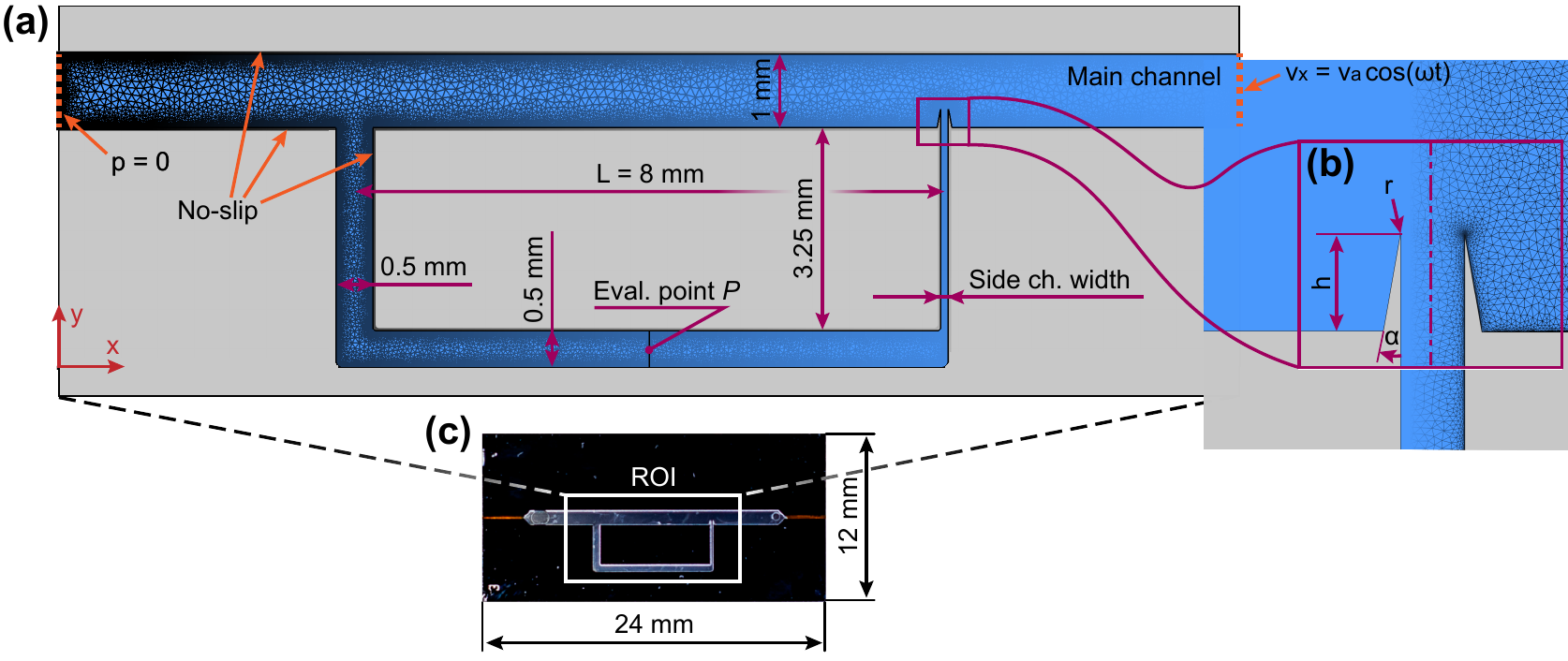}
 \caption{(a) Geometry and mesh used in the numerical simulations, with the description of geometrical parameters and boundary conditions applied in the numerical model. The blue region corresponds to the modeled fluid domain, whereas the grey area represents the solid structure of the device, which we assume to be rigid. (b) Detailed view of the geometry of sharp edges, with a refined mesh at the tips of sharp edges and at boundaries. The parameters that define the geometry of the sharp edge pair are the height of the sharp edges $h$, the angle of each sharp edge $\alpha$, and the rounding radius of the edges $r$. (c) Region of interest on an actual device that corresponds to the numerical model. The size of the chip in (c) is $12\times\SI{24}{\milli\meter}$.}
 \label{fig:fem_model}
\end{figure*}

\section*{Materials and methods}

%%% MATERIALS AND METHODS %%%

\subsection*{Devices and fabrication}
The lab-on-a-chip devices shown in Fig. \ref{fig:devices} were produced through cleanroom processing and feature a two-layered structure made of silicon and glass. The channel designs were first patterned on a silicon wafer ($500\pm\SI{10}{\micro\meter}$ thickness) through photolithography (resist:  S1828/S1818, Shipley, $4000 \, \mathrm{rpm}$; developer: AZ351B, Microchemicals) and then etched with an inductively coupled plasma deep reactive ion etching (ICP-DRIE) machine (Estrellas, Oxford instruments) to a depth of $\sim\SI{184}{\micro\meter}$.

Afterwards, a glass wafer ($\SI{500}{\micro\meter}$ thickness) was anodically bonded onto the silicon wafer. The wafer was then diced into individual $12\times\SI{24}{\milli\meter}$ chips with a wafer saw (DAD3221, Disco corporation). Fused silica capillaries ($164\pm\SI{6}{\micro\meter}$ outer diameter, $100\pm\SI{6}{\micro\meter}$ inner diameter, Molex) were inserted into the inlets and outlets of the chips and fixed with a two-component glue (5 Minute Epoxy, Devcon). A piezoelectric transducer (PZT) (length$\times$width$\times$thickness $=$ $10\times2\times\SI{1}{\milli\meter}$ or $16.75\times8.74\times\SI{17.4}{\milli\meter}$, Pz26, Meggitt Ferroperm) was glued to the backside of each device with a conductive epoxy (H20E, EPO-TEK). Copper cables ($\SI{0.15}{\milli\meter}$ diameter) were connected to the PZT using a conductive silver paste and glued to a device with instant glue for mechanical stability.

To fix the devices under the microscope, a chip-holder was designed and laser-cut from acrylic glass (PMMA), with top and bottom parts being held together by two bolts, as depicted in Fig. \ref{fig:devices}(a).

The devices were designed to have a $\SI{1}{\milli\meter}$ wide main channel across the length of the device, to which one or two loops of side channels  are connected, as visible in Fig. \ref{fig:devices}(b). An example of a single side channel loop device is shown in Fig. \ref{fig:fem_model}. The default width of side channels is $\SI{500}{\micro\meter}$, except for side channel segments that start with a pair of sharp edges, for which the width varies between $\SI{100}{\micro\meter}$ and $\SI{500}{\micro\meter}$. The width of this side channel segment is equal to the spacing between the pair of sharp edges, and is specified for each device alongside the corresponding results. All sharp edges in the study feature a $\SI{10}{\degree}$ apex angle, and are oriented such that one of the sides of the edge is aligned with a wall of the adjacent side channel. Measurements of the geometry of devices at the end of the production revealed the rounding radius of the sharp edges of $\sim \SI{0.9}{\micro\meter}$, the apex angle of a sharp edge of $\sim\SI{10.3}{\degree}$, the length of the protrusion of sharp edges into the main channel of $\sim \SI{275}{\micro\meter}$, and the deviation of channel widths in the order of $<\SI{1}{\percent}$.

\subsection*{Experimental setup}
For visual observation of phenomena within the channels of the lab-on-a-chip devices we used a microscope (Axioscope, Zeiss) with a blue LED (COP4-A, Thorlabs) and a high-speed camera (HiSpec1 2G Mono, Fastec Imaging). To visualize the flow field and demonstrate the acoustic forces, yeast cells (Coop Supermarket) and polystyrene (PS) beads ($5.19\pm \SI{0.14}{\micro\meter}$, microParticles GmbH) were used. Pressure pumps (Flow EZ, Fluigent) were connected to the inlet capillaries of a device, for supplying the particles/cells dispersed in water, as well as for re-dispersing the particles/cells within the chip in-between experiments. A function generator (Model DS345, Stanford Research Systems) and  an amplifier (HSA 4101, NF Corporation) were used to drive the PZT, and monitored through an oscilloscope (Model 9410, LeCroy).

Devices were characterized by measuring the admittance of the PZT with an impedance analyzer (Sine Phase Z-Check 16777k).$^\dag$ %The vibrational displacements and velocities were measured with a scaning laser vibrometer (MSA-500 Micro System Analyzer, Polytec).

The pumping velocity was estimated through manual particle tracking velocimetry (Fiji\cite{schindelin2012fiji}) of the PS particles. The flow rate was computed by assuming that the flow profile corresponds to the Poiseuille flow through a rectangular cross-section,\cite{bruus2008theoretical} and that the velocity measured in the middle of the channel corresponds to the maximal velocity of the flow profile. The trajectories of PS particles that were used for estimation of the acoustic pressure amplitude ($p_{\mathrm{a}}$) were obtained through a particle tracking velocimetry plugin TrackMate\cite{tinevez2017trackmate} in Fiji,\cite{schindelin2012fiji} and post-processed with a custom Matlab\cite{MATLAB:2019b} code (described in the supplemental material$^\dag$).

The time-averaged paths of PS particles through the system were obtained by applying the variance mode to a stack of several video frames in Adobe Photoshop.

%particle image velocimetry (PIVlab\cite{thielicke2014pivlab}) or

\subsection*{Numerical model}

The numerical model resembling the region of interest (ROI) of one of the devices, as indicated in Fig. \ref{fig:fem_model}(c), was based on a finite element method framework of COMSOL Multiphysics v. 5.6.\cite{comsol} The geometry with all the relevant dimensions, and the applied boundary conditions are shown in Fig. \ref{fig:fem_model}(a),(b). We only modeled the fluid domain in 2D, represented by the blue colored region in Fig. \ref{fig:fem_model}(a),(b), taking the walls as rigid boundaries. The material properties used for the modeling are given in Table \ref{tbl:mat_prop}.

\begin{table}[h]
\footnotesize
\centering
\bgroup
\def\arraystretch{1.1}
\begin{tabular}{l*{1}{c}r | l*{1}{c}r}
\hline \hline
   &\textbf{H2O} & \textbf{Unit} &  & \textbf{PS} & \textbf{Unit} \\
\hline
$\rho_0$&  $ 1000$ & $\SI{}{\kilo \gram \per \cubic \meter  }$ & $\rho_{\mathrm{obj}}$ & $1050$ & $\SI{}{\kilo \gram \per \cubic \meter  }$\\
$c_0$ &  $1481$ & $\SI{}{\meter \per \second  }$ & $c_{\mathrm{obj}}^{\mathrm{P}}$ & $2400$ & $\SI{}{\meter \per \second  }$\\
$\eta$ &  $1.002 $ & $\SI{}{\milli \pascal  \second  }$ & $c_{\mathrm{obj}}^{\mathrm{S}}$ & $1150$ & $\SI{}{\meter \per \second  }$\\
$\eta_{\mathrm{B}}$ &  $3.09 $ & $\SI{}{\milli \pascal  \second  }$ & $\Phi$ & $0.175$ & -\\
\hline \hline
\end{tabular}
\egroup
\caption{The material parameters for water (H2O) and polystyrene\cite{selfridge1985approximate} (PS). The solid elastic polystyrene is defined through the speed of primary and secondary waves, $c_{\mathrm{obj}}^{\mathrm{P}}$ and $c_{\mathrm{obj}}^{\mathrm{S}}$, respectively. The acoustic contrast factor $\Phi$ is computed using $\kappa_{\mathrm{obj}} = 1 / (\rho_{\mathrm{P}} [( c_{\mathrm{obj}}^{\mathrm{P}} )^2 - \frac{4}{3} ( c_{\mathrm{obj}}^{\mathrm{S}} )^2 ])$.}
\label{tbl:mat_prop}
\end{table}

The perturbation approach presented in the \textit{Theoretical background} section and in the supplemental material$^\dag$ was used for the numerical modeling, similarly to previous studies.\cite{muller2012numerical} The fluid domain is excited through boundary conditions specified in Fig. \ref{fig:fem_model}(a), by applying a zero acoustic pressure at the left-hand side boundary of the main channel, and an oscillatory velocity with an amplitude of $v_{\mathrm{a}}$ along the $x$-direction at the right-hand side boundary of the main channel. The rounding of the sharp edges is defined through the viscous boundary layer as $r = \delta / 10$, which ensures that the resulting streaming is independent of $r$, as demonstrated by \citet{zhang2020unveiling}. The detailed description of the numerical model is given in the supplemental material.$^\dag$

%If applicable: In line with previous studies, we use low frequency excitation in simulations to be able to analyse the sharp-edge driven pumping mechanism. This is in line with previous studies.

\section*{Results}

\begin{figure}[h]
 \centering
 \includegraphics{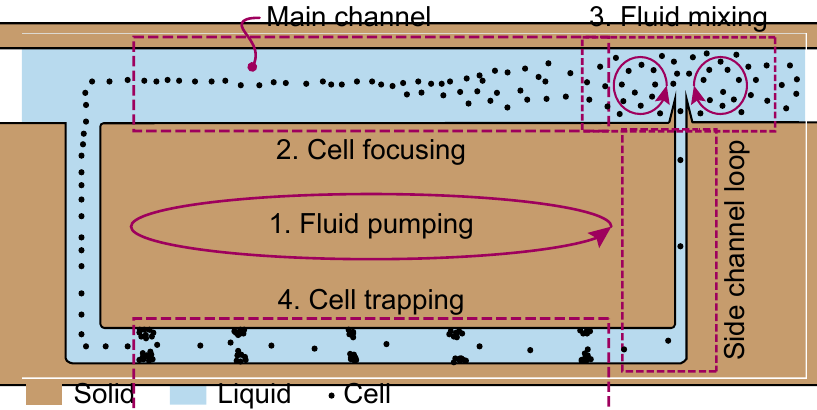}
 \caption{Overview of the four main functionalities of the device. (1.) Fluid pumping, (2.) cell/microparticle focusing, (3.) fluid mixing, and (4.) cell/microparticle trapping.}
 \label{fig:funct_overview}
\end{figure}

The investigated device features various functionalities that are graphically summarized in Fig. \ref{fig:funct_overview}. The two main being the pumping flow due to a single pair of sharp edges and the focusing of cells and microparticles due to the ARF in a standing acoustic wave. Other functionalities that are demonstrated in the manuscript are: trapping of cells with the localized pressure nodes in the side channels, and fluid mixing due to the strong acoustic streaming field near the two sharp edges. The results that demonstrate and characterize these functionalities are further discussed in the following sections.

\subsection*{Pumping and mixing}

\begin{figure*}
 \centering
 \includegraphics[scale=1]{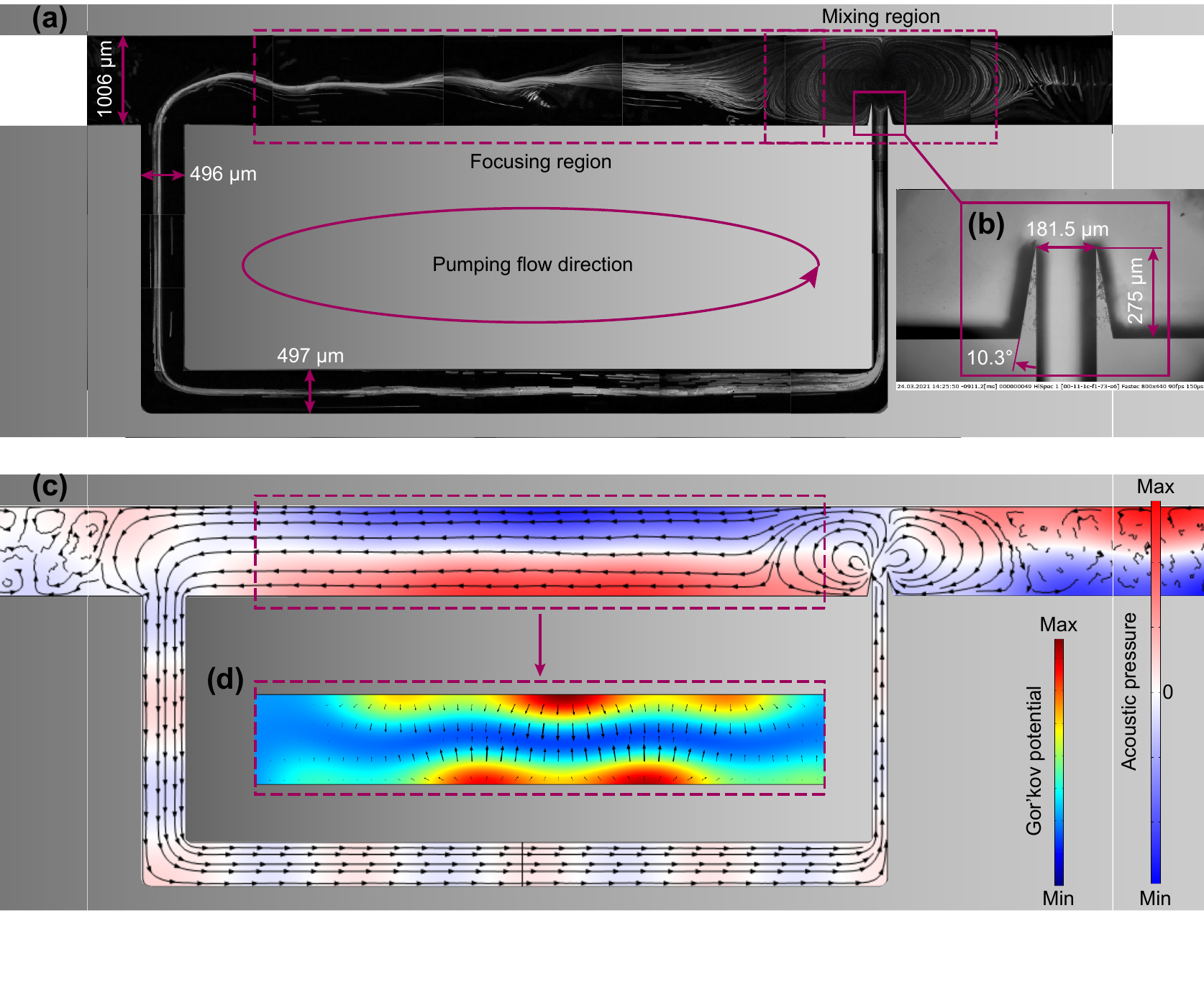}
 \caption{(a) Time-averaged paths of $\SI{5.19}{\micro\meter}$ polystyrene particles in the experimental device that is excited at $f=\SI{802}{\kilo\hertz}$ with $\SI{22}{\volt_{pp}}$.  The particles reveal the mixing around the two sharp edges, the focusing in the central region of the main channel, and the pumping through the main and side channel. Dimensions in (a) and (b) correspond to the measurements on the experimental device, yielding the spacing between the two edges of $\SI{181.5}{\micro\meter}$. (c) Numerically, we found a matching streaming field and an acoustic pressure field that yields the pressure node at the middle of the main channel, at $f=\SI{746}{\kilo\hertz}$. In the model, the spacing between the two edges was adjusted to $\SI{180}{\micro\meter}$, to match it to the spacing of the experimental device from (a). To avoid influence of effects near the ends of the main channel, the numerical model extends past the region that is shown here that matches the region from (a), observed in experiments. (d) The Gor'kov potential and the acoustic radiation force (arrows) further justify the focusing of the polystyrene particles at the pressure node. The time-averaged paths of PS particles through the system were obtained by applying the variance mode to a stack of several video frames in Adobe Photoshop. The direction of the experimentally observed flow near the sharp edges in (a) matches the direction of the numerically-predicted streaming in (c). }
 \label{fig:1}
\end{figure*}

\begin{figure}[h]
 \centering
 \includegraphics{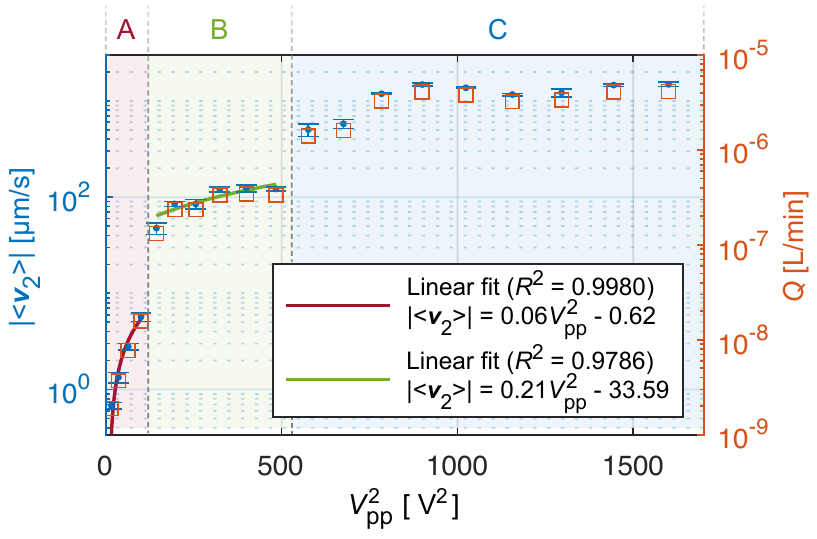}
 \caption{Velocity of the pumping flow in the middle of the side channel parallel to the main channel (equivalent to the evaluation point $P$ in Fig. \ref{fig:fem_model}), in dependence of the applied voltage squared. The device corresponds to the one in Fig. \ref{fig:1}, as well as the frequency, at $f=\SI{802}{\kilo\hertz}$. The graph is split into regions A: demonstrating a linear relationship between the streaming velocity and applied voltage squared, B: demonstrating the same linear relationship as in A but with a decreased slope, and C: a region where the streaming velocity cannot be increased by increasing the applied voltage squared, indicating a dominant influence of an unknown nonlinear physical mechanism.}
 \label{fig:regular_piezo_vpp_vel}
\end{figure}

\begin{figure}[h!]
 \centering
 \includegraphics[scale=1]{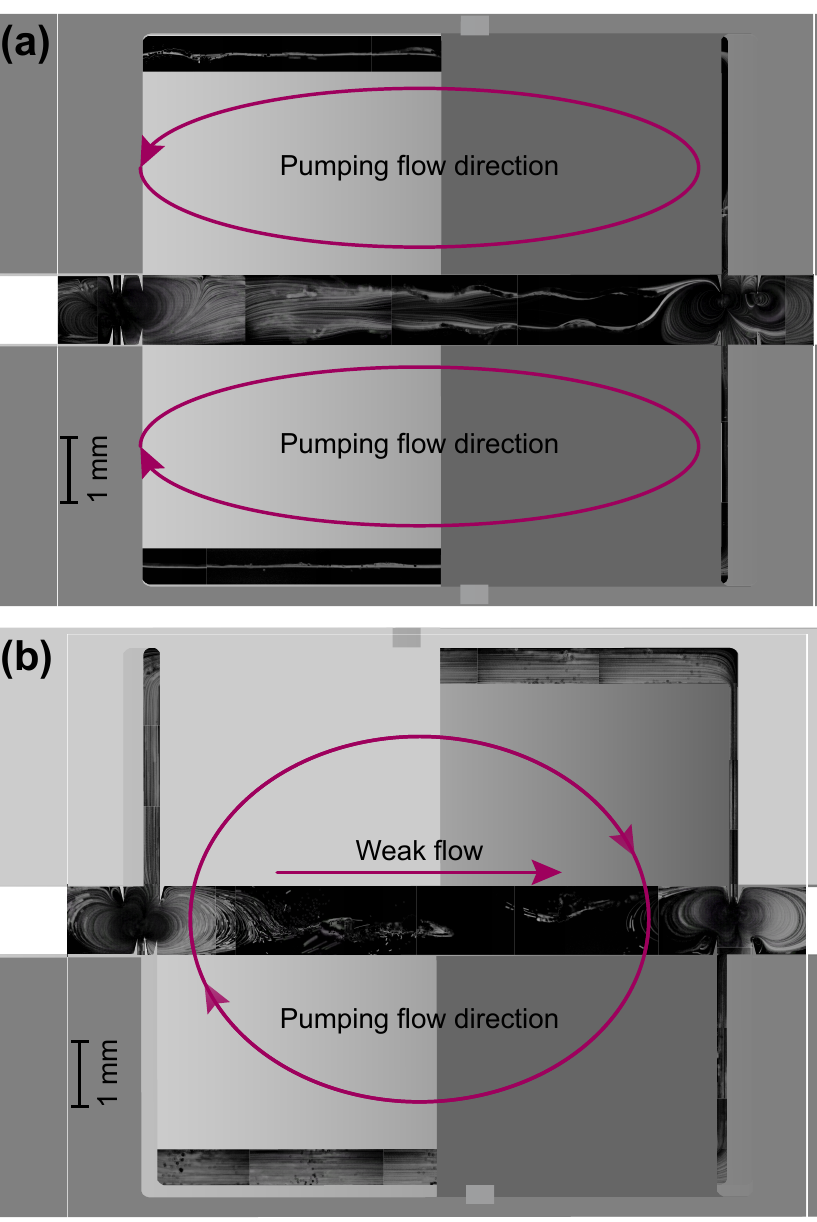}
 \caption{Time-averaged paths of $\SI{5.19}{\micro\meter}$ polystyrene particles in devices that feature two side channel loops with four pairs of sharp edges, with spacing between the two edges of $\SI{100}{\micro\meter}$ or $\SI{500}{\micro\meter}$. The width of the main channel of the two devices is $\SI{1}{\milli\meter}$. (a) Pairs of sharp edges with equal spacing between the two edges are opposing each other ($\SI{100}{\micro\meter}$ on the left and $\SI{500}{\micro\meter}$ on the right). The device was excited at $f=\SI{1.371}{\mega\hertz}$ with $V_{\mathrm{pp}} = \SI{30}{\volt}$. (b) Pairs of different spacing between the two edges are opposing each other ($\SI{100}{\micro\meter}$ bottom left and top right, $\SI{500}{\micro\meter}$ top left and bottom right). The device was excited at $f=\SI{780}{\kilo\hertz}$ with $V_{\mathrm{pp}} = \SI{6}{\volt}$. The arrows show the direction of the pumping flow, while the time-averaged paths of polystyrene particles reveal regions of focusing in (a) and a uniform flow in (b). The weak flow in the main channel in (b) is a result of a slight imbalance between the pumping strength of equally spaced sharp edge pairs, which is likely due to the spatial variability of the oscillatory velocity of the sharp edges relative to the liquid. The time-averaged paths of PS particles through the system were obtained by applying the variance mode to a stack of several video frames in Adobe Photoshop.}
 \label{fig:combined_c3_3and4}
\end{figure}

The pumping flow is demonstrated in Fig. \ref{fig:1}(a) through the time-averaged paths of $\SI{5.19}{\micro\meter}$ diameter PS particles in water. Near the two sharp edges, the particles follow the streamlines of the vortices that are typical around sharp edges,\cite{doinikov2020acoustic} and that are usable for mixing\cite{huang2013acoustofluidic,nama2016investigation,bachman2020acoustofluidic} at the micro scale. %The central region of the main channel is denoted as the focusing region, since the acoustic radiation force due to a $\lambda/2$ standing wave focuses the stream of PS particles to the middle of the channel.
The pumping flow in the counterclockwise direction, which results from the streaming around the two sharp edges,\cite{huang2014reliable} moves the particles from the mixing to the focusing region, and then back to the mixing region by pumping the particles through the side channel. The sharp edge angles of $\sim \SI{10}{\degree}$ are employed to maximize the streaming velocity, since it has been shown by \citet{zhang2020unveiling} and by \citet{doinikov2020acoustic} that the sharper the angle, the higher the streaming velocity at a fixed amplitude of the first-order excitation velocity. The video corresponding to Fig. \ref{fig:1}(a) is available in the supplemental material (\texttt{video\_1\_fig\_5}).$^\dag$
%Demonstrated and characterised functionality. The flow rate analysis comes here, as does the reference to the streaming patterns from experiments and simulations. Maybe join with the section on particle/cell focusing. Maybe estimate the pumping pressure, based on the amount of fluid being moved and the flow rates. Mention that others\cite{huang2014reliable} needed many sharp edge pairs to facilitate comparable flow rates.

The behaviour corresponding to the device from Fig. \ref{fig:1}(a) is analyzed with the help of the numerical model. The resulting acoustic pressure field is shown through colors in Fig. \ref{fig:1}(c), indicating that the pressure nodal plane appears near the middle of the main channel, at the frequency of $\SI{746}{\kilo\hertz}$. This frequency is lower than in an experimental setting (Fig. \ref{fig:1}a) due to simplifying the problem to two dimensions, and assuming rigid walls, both of which can influence the resonance frequency. The Gor'kov potential (color) and the ARF (arrows) in the focusing region are shown in Fig. \ref{fig:1}(d), and confirm that the PS particles are being forced to the middle of the channel. Furthermore, the simulation of the Eulerian acoustic streaming, the streamlines of which overlay the pressure field in Fig. \ref{fig:1}(c), indicates the pumping flow as well as the mixing flow close to the two edges. Qualitatively, our simplified two-dimensional numerical rigid-wall model that is based on the regular perturbation technique\cite{bruus2012perturbation,doinikov2020acoustic} captures the observed pumping and mixing phenomena very well.
However, the approach and the solution for the streaming velocity might be questionable in the proximity of the sharp tips, due to high velocities that can develop at those nearly-singular points.\cite{ovchinnikov2014acoustic} In order to improve the accuracy in that region, more sophisticated techniques would have to be applied, such as the direct numerical simulations (DNS)\cite{zhang2020unveiling,ovchinnikov2014acoustic} or the analytical approaches suitable for the so-called fast streaming.\cite{orosco2021unraveling} It is important to note that the region of concern near the sharp tips has a characteristic length in the order of the viscous boundary layer thickness $\delta = \sqrt{\eta / \pi \rho_0 f} = \SI{0.63}{\micro\meter}$ at $f=\SI{802}{\kilo\hertz}$ in water.

In the device from Fig. \ref{fig:1}, we measured the flow rate in dependence of the applied voltage squared, at $f = \SI{802}{\kilo\hertz}$, as shown in Fig. \ref{fig:regular_piezo_vpp_vel}. Specifically, we measured the velocity of the $\SI{5.19}{\micro\meter}$ PS particles in the pumping flow in the middle of the side channel that is parallel to the main channel, and transformed this velocity to the flow rate of a Poiseuille flow through the rectangular cross section of the channel.$^\dag$ At the applied voltages of below $V_{\mathrm{pp}} = \SI{10}{\volt}$, the good linear fit in the region ``A'' of Fig. \ref{fig:regular_piezo_vpp_vel} corresponds to the expected linear relation between the streaming velocity and the applied voltage squared.$^\dag$ However, between $V_{\mathrm{pp}} = \SI{10}{\volt}$ and $V_{\mathrm{pp}} = \SI{22}{\volt}$, in the region ``B'', the slope of the fitted linear curve is steeper, indicating a relative boost of the streaming velocity and a nonlinear dependence of the streaming velocity on the applied voltage squared. Interestingly, this is the opposite behaviour to that commonly observed in the streaming around sharp edges; for a single sharp edge, the slope of the increase in the streaming velocity with the increase of the applied voltage squared, would start decreasing after a certain threshold amplitude is reached.\cite{ovchinnikov2014acoustic,zhang2020unveiling}
There is another regime, at above $V_{\mathrm{pp}} = \SI{22}{\volt}$ - the region ``C'' in Fig. \ref{fig:regular_piezo_vpp_vel}, where the streaming velocity is more or less independent of the applied voltage. The observed nonlinear behaviour could indicate various physical mechanisms at play, such as the fluid turbulence, the maximal particle velocity threshold in the solid,\cite{singh2021investigation} or nonlinearities associated with the piezoelectric transducer and the conductive epoxy layer connecting the transducer to the device.

The source of the deviation from the linear relation between the streaming velocity magnitude and the applied voltage squared, in case of the sharp edge streaming, is not generally understood. As pointed out by \citet{ovchinnikov2014acoustic}, the nonlinearity could come from the increased influence of the third-order inertial term $\left( \left< \boldsymbol{v}_2 \right> \boldsymbol{\cdot} \boldsymbol{\nabla} \right) \boldsymbol{v}_1$ that is neglected in the second-order momentum conservation equation (eq. (S4) in the supplemental material$^\dag$); or from the interference of the streaming vortices with the surrounding geometry, according to \citet{zhang2020unveiling}.
Furthermore, the rounding radius of the sharp edges $r$ was measured to be $\sim \SI{0.9}{\micro\meter}$, which is comparable to the viscous boundary layer thickness in the investigated range ($\delta \approx \SI{0.56}{\micro\meter}$ at $f=\SI{1}{\mega\hertz}$ in water), and could be affecting the streaming flow. \citet{zhang2020unveiling} showed that the streaming velocity around a sharp edge decreases as $r$ increases, at $r > \delta$.

The flow rate in our devices reaches up to $\SI{4.1}{\micro\liter\per\minute}$ at $V_{\mathrm{pp}} = \SI{28}{\volt}$ and $f = \SI{802}{\kilo\hertz}$, which corresponds to the pumping pressure of $\sim \SI{12}{\pascal}$, approximated through the Hagen-Poiseuille law\cite{bruus2008theoretical}, considering the whole side channel loop. This is in the same order as the maximal flow rate and the pumping pressure generated in the previously reported PDMS acoustic sharp-edge micropump at $V_{\mathrm{pp}} = \SI{50}{\volt}$ and $f = \SI{6.5}{\kilo\hertz}$, namely $\sim \SI{8}{\micro\liter\per\minute}$ and $\sim \SI{76}{\pascal}$; however, we use a single pair of sharp edges, whereas the PDMS device features twenty sharp edges.

To demonstrate a wide range of operating frequencies that can be used to operate our micropumping mechanism, we used a larger PZT ($16.75\times8.74\times\SI{17.4}{\milli\meter}$) that has a resonance frequency in the thickness direction at $\sim \SI{80}{\kilo\hertz}$. The maximal flow rate observed at $f = \SI{78.26}{\kilo\hertz}$, for the same voltage range as in Fig. \ref{fig:regular_piezo_vpp_vel}, was $\SI{4.46}{\micro\liter\per\minute}$ at $V_{\mathrm{pp}} = \SI{35}{\volt}$. The corresponding video, showing the fluid pumping at $f = \SI{78.26}{\kilo\hertz}$ without the cell/microparticle focusing, is provided in the supplemental material$^\dag$ (\texttt{video\_2}).

Since the rate of the sharp edge pumping flow is controlled independently of the general flow through the whole device, which we control with the external pressure pump, we can fine-tune the flow rate in any of the side channel loops, and even bring it to stagnation, as shown in a video (\texttt{video\_3}) in the supplemental material.$^\dag$ This could potentially help increase the speed of operations like medium exchange or addition of a new sample, as the reduced flow in the side channel, while the PZT is excited, could impede the flushing of the trapped cells due to shear forces. The novelty of the device therefore lies in, but is not limited to, the programmable multifunctionality of the device, where each feature can either be used, in an arbitrary order, individually or simultaneously. %Fine-tuning the flow rate through the side channel loop, in combination with using the frequency at which trapping occurs in the side channel, could lead to size-selective or acoustic-contrast-selective up-concentration.

%Here, the general streaming patterns and particle traces are compared between numerical model and experiments. This gives a good idea that the model can be used to explain the experimental behaviour. Focus is on qualitative comparison of patterns (pressure/arf and streaming/pumping). Mention that stopping the flow in a side channel is possible + ref to a supplementary video. Show stitched images of the symmetric and asymmetric chips, or just one, of how the particles are focused + indicate the flow direction.

We also observed a dependence of the pumping flow on the spacing between the two edges. The pumping flows in the devices with two side channel loops in Fig. \ref{fig:combined_c3_3and4}, with two pairs of sharp edges per loop, indicate that the narrower $\SI{100}{\micro\meter}$ spacing between the sharp edges generates a stronger pumping flow than the $\SI{500}{\micro\meter}$ spacing, when other conditions (e.g. $V_{\mathrm{pp}}$, $f$, PZT placement, glue layer thickness, etc.) are kept constant.
For each loop in Fig. \ref{fig:combined_c3_3and4}, one of the sharp-edge pairs has a spacing of $\SI{100}{\micro\meter}$, while the spacing is $\SI{500}{\micro\meter}$ for the other. Both ends of a side channel loop should facilitate the pumping, as was verified through two additional devices featuring a single side channel loop (same design as in Fig. \ref{fig:1}a), with the spacing between the two edges of $\SI{100}{\micro\meter}$ and $\SI{500}{\micro\meter}$, respectively. However, based on the observed flow directions (Fig. \ref{fig:combined_c3_3and4}), the pumping from the sharp-edge pair with the narrower spacing prevails. %Due to the variability from device to device with regards to the placement and glueing of PZT, it is not possible to use multiple single side channel loop devices to quantitatively evaluate the influence of the spacing between the two edges on the generated pumping.

Experimental videos showcasing phenomena from Fig. \ref{fig:combined_c3_3and4} for PS particles (\texttt{video\_4\_fig\_7a} and \texttt{video\_5\_fig\_7b}) and for yeast cells (\texttt{video\_6} and \texttt{video\_7}) are available as part of the supplemental material.$^\dag$

The streaming vortices around the sharp edges that are analogous to ours, and that are responsible for the fluid mixing, have been investigated extensively,\cite{doinikov2020acoustic,zhang2019acoustic,zhang2020acoustic,zhang2020unveiling,ovchinnikov2014acoustic} which is why we omit a detailed analysis on the subject. %The mixing might, however, be a useful functionality for applications where, for example, two types of cells or two solutions need to be mixed together.
%\hl{(Add a bit more in-depth discussion on the mixing flow and its dependencies!)}

%The experimental results here should also highlight that the vortices become distinguishable when the side channel widens. Same with numerical results. The vortices could be analysed similarly to how Brunet et al. did it (using PIVlab). Vortex centers could also be compared (their positions), as the vorticity maps (check what that is).

\subsection*{Focusing and trapping of cells and microparticles}

%Demonstrated and characterised functionality. Here, the focusing patterns are discussed, providing the estimate on the acoustic energy density in the channel (estimate the $p_{\mathrm{a}}$ or $E_{\mathrm{ac}}$ from the out-of-loop segment of the video presented in Fig. \ref{fig:1}), and ability to focus yeast and polystyrene. Also discuss Gor'kov potential and the resulting acoustic radiation force from simulations. Can also be deactivated by exciting the PZT at an off-resonance frequency - example video is in the supplementary material (corresponding to one of the stitched videos - I think old C3 4). This could help the real-time imaging of the sample (e.g. cells), or could simply prevent cells and particles attaching to the walls of the device, which is a known problem for the long term operation of such acoustofluidic lab-on-a-chip devices.

Objects of positive acoustic contrast ($\Phi > 0$) are, in accordance with eq. (\ref{eq:Frad1D}), forced towards pressure nodes of a one-dimensional standing wave. This links the PS particles being focused towards the middle of the main channel in Fig. \ref{fig:1}(a), and the simulated acoustic pressure in Fig. \ref{fig:1}(c) indicating the pressure nodal line in the focusing region. The Gor'kov potential (color) and the ARF (arrows) in Fig. \ref{fig:1}(d), further confirm that the PS particles are forced towards the middle of the main channel.

The hard silicon walls of our devices provide a strong acoustic impedance mismatch with respect to water, $19.79 \cdot 10^6 \, \SI{}{\kilogram\per\metre\squared\per\second}$ compared to $1.49 \cdot 10^6 \, \SI{}{\kilogram\per\metre\squared\per\second}$, leading to the acoustically hard wall boundary condition for water at the solid-fluid interface ($\boldsymbol{v}_1 \approx \boldsymbol{0}$ at the interface) being a good model.\cite{lenshof2014building} This can result in a quasi one-dimensional standing wave between the walls of the main channel, featuring a $ n \lambda / 2 $ ultrasound resonance mode, where $n = 1, 2, 3, ...$ represents an individual resonance mode and $\lambda = c_0 / f$ the acoustic wavelength.\cite{bruus2012perturbation} For a fixed channel width, the resonance mode that is excited depends on the frequency of excitation: $\lambda / 2 $ mode, with its single pressure nodal plane, is visible at $f=\SI{802}{\kilo\hertz}$ in the $\SI{1}{\milli\meter}$ wide main channel in Fig. \ref{fig:1}(a), and at $f=\SI{1.371}{\mega\hertz}$ in the $\SI{0.5}{\milli\meter}$ wide side channels in Fig. \ref{fig:combined_c3_3and4}(a). The $\lambda$ resonance mode, featuring two pressure nodal lines, was observed at $f=\SI{1.371}{\mega\hertz}$ in the $\SI{1}{\milli\meter}$ wide main channel in Fig. \ref{fig:combined_c3_3and4}(a).

\begin{figure}[h]
 \centering
 \includegraphics{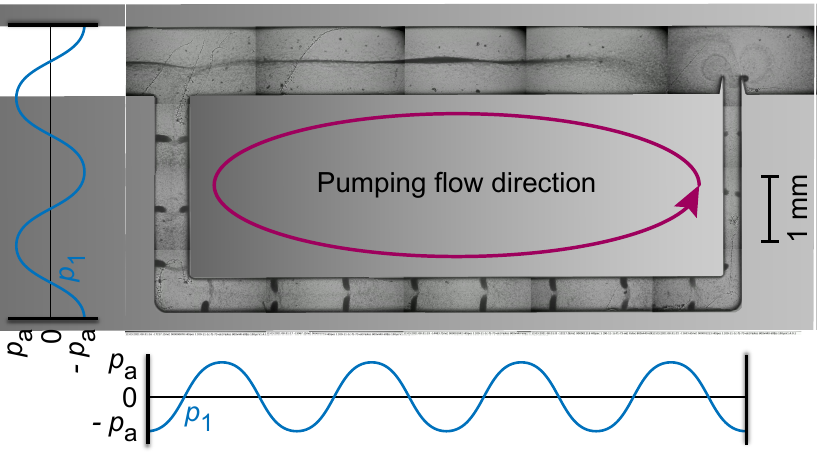}
 \caption{Trapping and focusing of yeast cells in water, demonstrated using a device featuring a $\SI{260}{\micro\meter}$ spacing between the two sharp edges, and is otherwise equal to the device from Fig. \ref{fig:1}, with $\SI{1}{\milli\meter}$ wide main channel. The image is a mosaic of several snapshots of channel segments, taken while the device was being excited at $f=\SI{714}{\kilo\hertz}$ with $V_{\mathrm{RMS}} = \SI{30}{\volt}$. The dark spots in the channel are the trapped clusters of yeast. The direction of the pumping flow is indicated by the arrow. The illustrated quasi-one-dimensional pressure fields along the side channels indicate the corresponding $n=4$ and $n=8$ resonance modes that focus and trap the yeast cells.}
 \label{fig:trapping}
\end{figure}

The acoustic pressure amplitude was measured in the region of the main channel outside the loop, where the mixing and pumping streaming flows were not evident (see the supplemental material$^\dag$ for details). The resulting average pressure amplitude $p_{\mathrm{a}}=\SI{1.2}{\mega\pascal}$ corresponds to $E_{\mathrm{ac}}=\SI{160}{\joule\per\meter\cubed}$, which is in the typical range for silicon-glass chips,\cite{bruus2012arf} considering the $V_{\mathrm{pp}} = \SI{22}{\volt}$ excitation.

In addition to the one-dimensional standing waves perpendicular to the channel mentioned so far, other resonance modes can be excited inside a device, yielding interesting trapping regions. Figure \ref{fig:trapping} shows clusters of trapped yeast cells in the side channel loop, showcasing the $n=4$ resonance mode in the two channels perpendicular to the main channel and the $n=8$ mode along the channel parallel to the main channel, while also yielding the $n=1$ mode across the main channel, with a single pressure nodal line. %Show a similar mode from simulations and explain that this could be beneficial for medium exchange and other applications like this. Additionally, cells/particles from multiple samples could be combined through this mechanism.
The trapped cells in Fig. \ref{fig:trapping} are disconnected across the channel width in the side channel loop due to the pumping flow that features the highest velocity in the middle of the channel cross-section and obstructs the trapping.

Such trapping capabilities, in combination with the flow control through the side channel, could improve the efficiency of the cell isolation process\cite{jiang2021rapid} or the medium exchange.\cite{gerlt2021acoustofluidic} The acoustic radiation force that is the mechanism behind the trapping and focusing in our device, scales with the volume of a cell, whereas the drag force from the acoustic streaming scales with the radius of the cell. Consequently, fine-tuning the flow rate through the side channel loop and the strength of the trapping, by modifying the voltage applied to the PZT and the external pressure difference applied to the main channel, could lead to size-selective separation of cells, with a programmable size threshold.

\section*{Conclusions}

% Moving fluid around on the micro scale is generally a hard task. There is a field of research and development dealing with such technology\cite{wang2018micropumps,dereshgi2021piezoelectric}.
The presented microfluidic chip offers a platform that is especially relevant for applications where multiple functions are needed within the same chip, particularly pumping, mixing, and focusing or trapping of cells and microparticles. We demonstrated a continuous pumping in the range of excitation frequencies from $\SI{80}{\kilo\hertz}$ to above $\SI{1}{\mega\hertz}$, with the corresponding programmable flow rates of up to $\sim \SI{4}{\micro\liter\per\minute}$ ($\sim \SI{12}{\pascal}$ pumping pressure). The pumping is always paired with the mixing vortices near the sharp edges that drive the pumping flow. We also demonstrated the focusing and trapping capabilities of our device by predicting and observing the behavior of polystyrene microparticles and yeast cells in combination with different resonance modes of the fluidic channels.

Our silicon-glass device is chemically resistant, solving one of the main problems of acoustic microreactors for chemical engineering.\cite{CHEN2021133258} The pumping and strong mixing at the sharp edges could provide a tool for cell disruption, similar to what \citet{hughes1962cell} demonstrated, but with the ability to disrupt larger samples in a very controlled manner. The unique precisely programmable multifunctionality that the device can provide, could also be beneficial in studying the dynamics of active matter, such as bacteria\cite{gutierrez2018induced} or Janus particles.\cite{takatori2016acoustic} Furthermore, it has been previously demonstrated that imposing a standing acoustic wave can significantly improve the sensitivity of particle-based immunoassays.\cite{wiklund2013acoustofluidics} The combination of mixing, focusing, and pumping that our device provides, could lead to an even higher sensitivity of these methods.

In the future, the device design will be further optimized based on the needs of specific applications. The process will be aided by the presented numerical model that is capable of predicting the acoustofluidic phenomena inside such lab-on-a-chip devices.
Furthermore, the thin film bulk acoustic wave (TFBAW) technology\cite{reichert2018thin} could be applied to mitigate the variability in production, and to potentially improve the efficiency of the devices. This would also allow for a more in-depth experimental analysis of the influence of various geometrical and excitation parameters on the observed phenomena, further aiding the optimization of the device design.

\section*{Conflicts of interest}
There are no conflicts to declare.

% \section*{Acknowledgements}
% The Acknowledgements come at the end of an article after Conflicts of interest and before the Notes and references.

%%%END OF MAIN TEXT%%%

%\section*{Supplementary material}

%\input{sections/supplementary}

%The \balance command can be used to balance the columns on the final page if desired. It should be placed anywhere within the first column of the last page.

\balance

%If notes are included in your references you can change the title from 'References' to 'Notes and references' using the following command:
%\renewcommand\refname{Notes and references}

%%%REFERENCES%%%
\bibliography{rsc} %You need to replace "rsc" on this line with the name of your .bib file

\providecommand*{\mcitethebibliography}{\thebibliography}
\csname @ifundefined\endcsname{endmcitethebibliography}
{\let\endmcitethebibliography\endthebibliography}{}
\begin{mcitethebibliography}{59}
\providecommand*{\natexlab}[1]{#1}
\providecommand*{\mciteSetBstSublistMode}[1]{}
\providecommand*{\mciteSetBstMaxWidthForm}[2]{}
\providecommand*{\mciteBstWouldAddEndPuncttrue}
  {\def\EndOfBibitem{\unskip.}}
\providecommand*{\mciteBstWouldAddEndPunctfalse}
  {\let\EndOfBibitem\relax}
\providecommand*{\mciteSetBstMidEndSepPunct}[3]{}
\providecommand*{\mciteSetBstSublistLabelBeginEnd}[3]{}
\providecommand*{\EndOfBibitem}{}
\mciteSetBstSublistMode{f}
\mciteSetBstMaxWidthForm{subitem}
{(\emph{\alph{mcitesubitemcount}})}
\mciteSetBstSublistLabelBeginEnd{\mcitemaxwidthsubitemform\space}
{\relax}{\relax}

\bibitem[Wang and Fu(2018)]{wang2018micropumps}
Y.-N. Wang and L.-M. Fu, \emph{Microelectronic Engineering}, 2018,
  \textbf{195}, 121--138\relax
\mciteBstWouldAddEndPuncttrue
\mciteSetBstMidEndSepPunct{\mcitedefaultmidpunct}
{\mcitedefaultendpunct}{\mcitedefaultseppunct}\relax
\EndOfBibitem
\bibitem[Byun \emph{et~al.}(2014)Byun, Abi-Samra, Cho, and
  Takayama]{byun2014pumps}
C.~K. Byun, K.~Abi-Samra, Y.-K. Cho and S.~Takayama, \emph{Electrophoresis},
  2014, \textbf{35}, 245--257\relax
\mciteBstWouldAddEndPuncttrue
\mciteSetBstMidEndSepPunct{\mcitedefaultmidpunct}
{\mcitedefaultendpunct}{\mcitedefaultseppunct}\relax
\EndOfBibitem
\bibitem[Dereshgi \emph{et~al.}(2021)Dereshgi, Dal, and
  Yildiz]{dereshgi2021piezoelectric}
H.~A. Dereshgi, H.~Dal and M.~Z. Yildiz, \emph{Microsystem Technologies}, 2021,
   1--29\relax
\mciteBstWouldAddEndPuncttrue
\mciteSetBstMidEndSepPunct{\mcitedefaultmidpunct}
{\mcitedefaultendpunct}{\mcitedefaultseppunct}\relax
\EndOfBibitem
\bibitem[Kim \emph{et~al.}(2014)Kim, Astle, Najafi, Bernal, and
  Washabaugh]{kim2014integrated}
H.~Kim, A.~A. Astle, K.~Najafi, L.~P. Bernal and P.~D. Washabaugh,
  \emph{Journal of Microelectromechanical Systems}, 2014, \textbf{24},
  192--206\relax
\mciteBstWouldAddEndPuncttrue
\mciteSetBstMidEndSepPunct{\mcitedefaultmidpunct}
{\mcitedefaultendpunct}{\mcitedefaultseppunct}\relax
\EndOfBibitem
\bibitem[Rusli \emph{et~al.}(2018)Rusli, Chee, Arsat, Lau, and
  Leow]{rusli2018electromagnetic}
M.~Rusli, P.~S. Chee, R.~Arsat, K.~X. Lau and P.~L. Leow, \emph{Sensors and
  Actuators A: Physical}, 2018, \textbf{282}, 17--27\relax
\mciteBstWouldAddEndPuncttrue
\mciteSetBstMidEndSepPunct{\mcitedefaultmidpunct}
{\mcitedefaultendpunct}{\mcitedefaultseppunct}\relax
\EndOfBibitem
\bibitem[Hamid \emph{et~al.}(2017)Hamid, Majlis, Yunas, Syafeeza, Wong, and
  Ibrahim]{hamid2017stack}
N.~A. Hamid, B.~Y. Majlis, J.~Yunas, A.~Syafeeza, Y.~C. Wong and M.~Ibrahim,
  \emph{Microsystem Technologies}, 2017, \textbf{23}, 4037--4043\relax
\mciteBstWouldAddEndPuncttrue
\mciteSetBstMidEndSepPunct{\mcitedefaultmidpunct}
{\mcitedefaultendpunct}{\mcitedefaultseppunct}\relax
\EndOfBibitem
\bibitem[Oskooei and G{\"u}nther(2015)]{oskooei2015bubble}
A.~Oskooei and A.~G{\"u}nther, \emph{Lab on a Chip}, 2015, \textbf{15},
  2842--2853\relax
\mciteBstWouldAddEndPuncttrue
\mciteSetBstMidEndSepPunct{\mcitedefaultmidpunct}
{\mcitedefaultendpunct}{\mcitedefaultseppunct}\relax
\EndOfBibitem
\bibitem[Khoshmanesh \emph{et~al.}(2017)Khoshmanesh, Tang, Zhu, Schaefer,
  Mitchell, Kalantar-Zadeh, and Dickey]{khoshmanesh2017liquid}
K.~Khoshmanesh, S.-Y. Tang, J.~Y. Zhu, S.~Schaefer, A.~Mitchell,
  K.~Kalantar-Zadeh and M.~D. Dickey, \emph{Lab on a Chip}, 2017, \textbf{17},
  974--993\relax
\mciteBstWouldAddEndPuncttrue
\mciteSetBstMidEndSepPunct{\mcitedefaultmidpunct}
{\mcitedefaultendpunct}{\mcitedefaultseppunct}\relax
\EndOfBibitem
\bibitem[Huang \emph{et~al.}(2014)Huang, Nama, Mao, Li, Rufo, Chen, Xie, Wei,
  Wang, and Huang]{huang2014reliable}
P.-H. Huang, N.~Nama, Z.~Mao, P.~Li, J.~Rufo, Y.~Chen, Y.~Xie, C.-H. Wei,
  L.~Wang and T.~J. Huang, \emph{Lab on a Chip}, 2014, \textbf{14},
  4319--4323\relax
\mciteBstWouldAddEndPuncttrue
\mciteSetBstMidEndSepPunct{\mcitedefaultmidpunct}
{\mcitedefaultendpunct}{\mcitedefaultseppunct}\relax
\EndOfBibitem
\bibitem[Wang \emph{et~al.}(2010)Wang, Huang, and Yang]{wang2010valveless}
S.~Wang, X.~Huang and C.~Yang, \emph{Microfluidics and Nanofluidics}, 2010,
  \textbf{8}, 549--555\relax
\mciteBstWouldAddEndPuncttrue
\mciteSetBstMidEndSepPunct{\mcitedefaultmidpunct}
{\mcitedefaultendpunct}{\mcitedefaultseppunct}\relax
\EndOfBibitem
\bibitem[Ryu \emph{et~al.}(2010)Ryu, Chung, and Cho]{ryu2010micropumping}
K.~Ryu, S.~K. Chung and S.~K. Cho, \emph{JALA: Journal of the Association for
  Laboratory Automation}, 2010, \textbf{15}, 163--171\relax
\mciteBstWouldAddEndPuncttrue
\mciteSetBstMidEndSepPunct{\mcitedefaultmidpunct}
{\mcitedefaultendpunct}{\mcitedefaultseppunct}\relax
\EndOfBibitem
\bibitem[Nabavi \emph{et~al.}(2008)Nabavi, Siddiqui, and
  Dargahi]{nabavi2008analysis}
M.~Nabavi, K.~Siddiqui and J.~Dargahi, \emph{Physics of Fluids}, 2008,
  \textbf{20}, 126101\relax
\mciteBstWouldAddEndPuncttrue
\mciteSetBstMidEndSepPunct{\mcitedefaultmidpunct}
{\mcitedefaultendpunct}{\mcitedefaultseppunct}\relax
\EndOfBibitem
\bibitem[Ozcelik and Aslan(2021)]{ozcelik2021practical}
A.~Ozcelik and Z.~Aslan, \emph{Microfluidics and Nanofluidics}, 2021,
  \textbf{25}, 1--10\relax
\mciteBstWouldAddEndPuncttrue
\mciteSetBstMidEndSepPunct{\mcitedefaultmidpunct}
{\mcitedefaultendpunct}{\mcitedefaultseppunct}\relax
\EndOfBibitem
\bibitem[Lin \emph{et~al.}(2019)Lin, Gao, Wu, Zhou, Chung, Caraveo, and
  Xu]{lin2019acoustofluidic}
Y.~Lin, Y.~Gao, M.~Wu, R.~Zhou, D.~Chung, G.~Caraveo and J.~Xu, \emph{Lab on a
  Chip}, 2019, \textbf{19}, 3045--3053\relax
\mciteBstWouldAddEndPuncttrue
\mciteSetBstMidEndSepPunct{\mcitedefaultmidpunct}
{\mcitedefaultendpunct}{\mcitedefaultseppunct}\relax
\EndOfBibitem
\bibitem[Jiang \emph{et~al.}(2021)Jiang, Agrawal, Aghaamoo, Parajuli, Agrawal,
  and Lee]{jiang2021rapid}
R.~Jiang, S.~Agrawal, M.~Aghaamoo, R.~Parajuli, A.~Agrawal and A.~P. Lee,
  \emph{Lab on a Chip}, 2021, \textbf{21}, 875--887\relax
\mciteBstWouldAddEndPuncttrue
\mciteSetBstMidEndSepPunct{\mcitedefaultmidpunct}
{\mcitedefaultendpunct}{\mcitedefaultseppunct}\relax
\EndOfBibitem
\bibitem[Lighthill(1978)]{lighthill1978acoustic}
J.~Lighthill, \emph{Journal of Sound and Vibration}, 1978, \textbf{61},
  391--418\relax
\mciteBstWouldAddEndPuncttrue
\mciteSetBstMidEndSepPunct{\mcitedefaultmidpunct}
{\mcitedefaultendpunct}{\mcitedefaultseppunct}\relax
\EndOfBibitem
\bibitem[Rayleigh(1884)]{rayleigh1884circulation}
L.~Rayleigh, \emph{Philosophical Transactions of the Royal Society of London},
  1884, \textbf{175}, 1--21\relax
\mciteBstWouldAddEndPuncttrue
\mciteSetBstMidEndSepPunct{\mcitedefaultmidpunct}
{\mcitedefaultendpunct}{\mcitedefaultseppunct}\relax
\EndOfBibitem
\bibitem[Schlichting(1932)]{schlichting1932berechnung}
H.~Schlichting, \emph{Phys. z.}, 1932, \textbf{33}, 327--335\relax
\mciteBstWouldAddEndPuncttrue
\mciteSetBstMidEndSepPunct{\mcitedefaultmidpunct}
{\mcitedefaultendpunct}{\mcitedefaultseppunct}\relax
\EndOfBibitem
\bibitem[Eckart(1948)]{eckart1948vortices}
C.~Eckart, \emph{Physical Review}, 1948, \textbf{73}, 68\relax
\mciteBstWouldAddEndPuncttrue
\mciteSetBstMidEndSepPunct{\mcitedefaultmidpunct}
{\mcitedefaultendpunct}{\mcitedefaultseppunct}\relax
\EndOfBibitem
\bibitem[Zhang \emph{et~al.}(2021)Zhang, Horesh, Manor, and
  Friend]{zhang2021powerful}
N.~Zhang, A.~Horesh, O.~Manor and J.~Friend, \emph{Physical Review Letters},
  2021, \textbf{126}, 164502\relax
\mciteBstWouldAddEndPuncttrue
\mciteSetBstMidEndSepPunct{\mcitedefaultmidpunct}
{\mcitedefaultendpunct}{\mcitedefaultseppunct}\relax
\EndOfBibitem
\bibitem[Hughes and Nyborg(1962)]{hughes1962cell}
D.~Hughes and W.~Nyborg, \emph{Science}, 1962, \textbf{138}, 108--114\relax
\mciteBstWouldAddEndPuncttrue
\mciteSetBstMidEndSepPunct{\mcitedefaultmidpunct}
{\mcitedefaultendpunct}{\mcitedefaultseppunct}\relax
\EndOfBibitem
\bibitem[Zhang \emph{et~al.}(2019)Zhang, Guo, Brunet, Costalonga, and
  Royon]{zhang2019acoustic}
C.~Zhang, X.~Guo, P.~Brunet, M.~Costalonga and L.~Royon, \emph{Microfluidics
  and Nanofluidics}, 2019, \textbf{23}, 1--15\relax
\mciteBstWouldAddEndPuncttrue
\mciteSetBstMidEndSepPunct{\mcitedefaultmidpunct}
{\mcitedefaultendpunct}{\mcitedefaultseppunct}\relax
\EndOfBibitem
\bibitem[Zhang \emph{et~al.}(2020)Zhang, Guo, Royon, and
  Brunet]{zhang2020acoustic}
C.~Zhang, X.~Guo, L.~Royon and P.~Brunet, \emph{Micromachines}, 2020,
  \textbf{11}, 607\relax
\mciteBstWouldAddEndPuncttrue
\mciteSetBstMidEndSepPunct{\mcitedefaultmidpunct}
{\mcitedefaultendpunct}{\mcitedefaultseppunct}\relax
\EndOfBibitem
\bibitem[Zhang \emph{et~al.}(2020)Zhang, Guo, Royon, and
  Brunet]{zhang2020unveiling}
C.~Zhang, X.~Guo, L.~Royon and P.~Brunet, \emph{Physical Review E}, 2020,
  \textbf{102}, 043110\relax
\mciteBstWouldAddEndPuncttrue
\mciteSetBstMidEndSepPunct{\mcitedefaultmidpunct}
{\mcitedefaultendpunct}{\mcitedefaultseppunct}\relax
\EndOfBibitem
\bibitem[Huang \emph{et~al.}(2013)Huang, Xie, Ahmed, Rufo, Nama, Chen, Chan,
  and Huang]{huang2013acoustofluidic}
P.-H. Huang, Y.~Xie, D.~Ahmed, J.~Rufo, N.~Nama, Y.~Chen, C.~Y. Chan and T.~J.
  Huang, \emph{Lab on a Chip}, 2013, \textbf{13}, 3847--3852\relax
\mciteBstWouldAddEndPuncttrue
\mciteSetBstMidEndSepPunct{\mcitedefaultmidpunct}
{\mcitedefaultendpunct}{\mcitedefaultseppunct}\relax
\EndOfBibitem
\bibitem[Nama \emph{et~al.}(2016)Nama, Huang, Huang, and
  Costanzo]{nama2016investigation}
N.~Nama, P.-H. Huang, T.~J. Huang and F.~Costanzo, \emph{Biomicrofluidics},
  2016, \textbf{10}, 024124\relax
\mciteBstWouldAddEndPuncttrue
\mciteSetBstMidEndSepPunct{\mcitedefaultmidpunct}
{\mcitedefaultendpunct}{\mcitedefaultseppunct}\relax
\EndOfBibitem
\bibitem[Bachman \emph{et~al.}(2020)Bachman, Chen, Rufo, Zhao, Yang, Tian,
  Nama, Huang, and Huang]{bachman2020acoustofluidic}
H.~Bachman, C.~Chen, J.~Rufo, S.~Zhao, S.~Yang, Z.~Tian, N.~Nama, P.-H. Huang
  and T.~J. Huang, \emph{Lab on a Chip}, 2020, \textbf{20}, 1238--1248\relax
\mciteBstWouldAddEndPuncttrue
\mciteSetBstMidEndSepPunct{\mcitedefaultmidpunct}
{\mcitedefaultendpunct}{\mcitedefaultseppunct}\relax
\EndOfBibitem
\bibitem[Doinikov \emph{et~al.}(2020)Doinikov, Gerlt, Pavlic, and
  Dual]{doinikov2020acoustic}
A.~A. Doinikov, M.~S. Gerlt, A.~Pavlic and J.~Dual, \emph{Microfluidics and
  Nanofluidics}, 2020, \textbf{24}, 1--13\relax
\mciteBstWouldAddEndPuncttrue
\mciteSetBstMidEndSepPunct{\mcitedefaultmidpunct}
{\mcitedefaultendpunct}{\mcitedefaultseppunct}\relax
\EndOfBibitem
\bibitem[Leibacher \emph{et~al.}(2015)Leibacher, Hahn, and
  Dual]{leibacher2015acoustophoretic}
I.~Leibacher, P.~Hahn and J.~Dual, \emph{Microfluidics and Nanofluidics}, 2015,
  \textbf{19}, 923--933\relax
\mciteBstWouldAddEndPuncttrue
\mciteSetBstMidEndSepPunct{\mcitedefaultmidpunct}
{\mcitedefaultendpunct}{\mcitedefaultseppunct}\relax
\EndOfBibitem
\bibitem[Doinikov \emph{et~al.}(2020)Doinikov, Gerlt, and
  Dual]{doinikov2020arf}
A.~A. Doinikov, M.~S. Gerlt and J.~Dual, \emph{Physical Review Letters}, 2020,
  \textbf{124}, 154501\relax
\mciteBstWouldAddEndPuncttrue
\mciteSetBstMidEndSepPunct{\mcitedefaultmidpunct}
{\mcitedefaultendpunct}{\mcitedefaultseppunct}\relax
\EndOfBibitem
\bibitem[Ovchinnikov \emph{et~al.}(2014)Ovchinnikov, Zhou, and
  Yalamanchili]{ovchinnikov2014acoustic}
M.~Ovchinnikov, J.~Zhou and S.~Yalamanchili, \emph{The Journal of the
  Acoustical Society of America}, 2014, \textbf{136}, 22--29\relax
\mciteBstWouldAddEndPuncttrue
\mciteSetBstMidEndSepPunct{\mcitedefaultmidpunct}
{\mcitedefaultendpunct}{\mcitedefaultseppunct}\relax
\EndOfBibitem
\bibitem[Evander and Nilsson(2012)]{evander2012acoustofluidics}
M.~Evander and J.~Nilsson, \emph{Lab on a Chip}, 2012, \textbf{12},
  4667--4676\relax
\mciteBstWouldAddEndPuncttrue
\mciteSetBstMidEndSepPunct{\mcitedefaultmidpunct}
{\mcitedefaultendpunct}{\mcitedefaultseppunct}\relax
\EndOfBibitem
\bibitem[Wu \emph{et~al.}(2019)Wu, Ozcelik, Rufo, Wang, Fang, and
  Huang]{wu2019acoustofluidic}
M.~Wu, A.~Ozcelik, J.~Rufo, Z.~Wang, R.~Fang and T.~J. Huang,
  \emph{Microsystems \& Nanoengineering}, 2019, \textbf{5}, 1--18\relax
\mciteBstWouldAddEndPuncttrue
\mciteSetBstMidEndSepPunct{\mcitedefaultmidpunct}
{\mcitedefaultendpunct}{\mcitedefaultseppunct}\relax
\EndOfBibitem
\bibitem[Leibacher \emph{et~al.}(2015)Leibacher, Reichert, and
  Dual]{leibacher2015microfluidic}
I.~Leibacher, P.~Reichert and J.~Dual, \emph{Lab on a Chip}, 2015, \textbf{15},
  2896--2905\relax
\mciteBstWouldAddEndPuncttrue
\mciteSetBstMidEndSepPunct{\mcitedefaultmidpunct}
{\mcitedefaultendpunct}{\mcitedefaultseppunct}\relax
\EndOfBibitem
\bibitem[Bruus \emph{et~al.}(2011)Bruus, Dual, Hawkes, Hill, Laurell, Nilsson,
  Radel, Sadhal, and Wiklund]{bruus2011forthcoming}
H.~Bruus, J.~Dual, J.~Hawkes, M.~Hill, T.~Laurell, J.~Nilsson, S.~Radel,
  S.~Sadhal and M.~Wiklund, \emph{Lab on a Chip}, 2011, \textbf{11},
  3579--3580\relax
\mciteBstWouldAddEndPuncttrue
\mciteSetBstMidEndSepPunct{\mcitedefaultmidpunct}
{\mcitedefaultendpunct}{\mcitedefaultseppunct}\relax
\EndOfBibitem
\bibitem[{\"O}zer and {\c{C}}etin(2021)]{ozer2021extended}
M.~B. {\"O}zer and B.~{\c{C}}etin, \emph{The Journal of the Acoustical Society
  of America}, 2021, \textbf{149}, 2802--2812\relax
\mciteBstWouldAddEndPuncttrue
\mciteSetBstMidEndSepPunct{\mcitedefaultmidpunct}
{\mcitedefaultendpunct}{\mcitedefaultseppunct}\relax
\EndOfBibitem
\bibitem[Moiseyenko and Bruus(2019)]{moiseyenko2019whole}
R.~P. Moiseyenko and H.~Bruus, \emph{Physical Review Applied}, 2019,
  \textbf{11}, 014014\relax
\mciteBstWouldAddEndPuncttrue
\mciteSetBstMidEndSepPunct{\mcitedefaultmidpunct}
{\mcitedefaultendpunct}{\mcitedefaultseppunct}\relax
\EndOfBibitem
\bibitem[Chen \emph{et~al.}(2021)Chen, Pei, Zhao, Zhang, Wei, and
  Hao]{CHEN2021133258}
Z.~Chen, Z.~Pei, X.~Zhao, J.~Zhang, J.~Wei and N.~Hao, \emph{Chemical
  Engineering Journal}, 2021,  133258\relax
\mciteBstWouldAddEndPuncttrue
\mciteSetBstMidEndSepPunct{\mcitedefaultmidpunct}
{\mcitedefaultendpunct}{\mcitedefaultseppunct}\relax
\EndOfBibitem
\bibitem[Wiklund(2012)]{wiklund2012acoustofluidics}
M.~Wiklund, \emph{Lab on a Chip}, 2012, \textbf{12}, 2018--2028\relax
\mciteBstWouldAddEndPuncttrue
\mciteSetBstMidEndSepPunct{\mcitedefaultmidpunct}
{\mcitedefaultendpunct}{\mcitedefaultseppunct}\relax
\EndOfBibitem
\bibitem[Wiklund \emph{et~al.}(2013)Wiklund, Radel, and
  Hawkes]{wiklund2013acoustofluidics}
M.~Wiklund, S.~Radel and J.~J. Hawkes, \emph{Lab on a Chip}, 2013, \textbf{13},
  25--39\relax
\mciteBstWouldAddEndPuncttrue
\mciteSetBstMidEndSepPunct{\mcitedefaultmidpunct}
{\mcitedefaultendpunct}{\mcitedefaultseppunct}\relax
\EndOfBibitem
\bibitem[Gerlt \emph{et~al.}(2021)Gerlt, Ruppen, Leuthner, Panke, and
  Dual]{gerlt2021acoustofluidic}
M.~Gerlt, P.~Ruppen, M.~Leuthner, S.~Panke and J.~Dual, 2021\relax
\mciteBstWouldAddEndPuncttrue
\mciteSetBstMidEndSepPunct{\mcitedefaultmidpunct}
{\mcitedefaultendpunct}{\mcitedefaultseppunct}\relax
\EndOfBibitem
\bibitem[Bruus(2012)]{bruus2012perturbation}
H.~Bruus, \emph{Lab on a Chip}, 2012, \textbf{12}, 20--28\relax
\mciteBstWouldAddEndPuncttrue
\mciteSetBstMidEndSepPunct{\mcitedefaultmidpunct}
{\mcitedefaultendpunct}{\mcitedefaultseppunct}\relax
\EndOfBibitem
\bibitem[Gor'kov(1962)]{gor1962forces}
L.~P. Gor'kov, Sov. Phys. Dokl., 1962, pp. 773--775\relax
\mciteBstWouldAddEndPuncttrue
\mciteSetBstMidEndSepPunct{\mcitedefaultmidpunct}
{\mcitedefaultendpunct}{\mcitedefaultseppunct}\relax
\EndOfBibitem
\bibitem[Baasch and Dual(2020)]{baasch2020acoustic}
T.~Baasch and J.~Dual, \emph{Physical Review Applied}, 2020, \textbf{14},
  024052\relax
\mciteBstWouldAddEndPuncttrue
\mciteSetBstMidEndSepPunct{\mcitedefaultmidpunct}
{\mcitedefaultendpunct}{\mcitedefaultseppunct}\relax
\EndOfBibitem
\bibitem[Yosioka and Kawasima(1955)]{yosioka1955acoustic}
K.~Yosioka and Y.~Kawasima, \emph{Acta Acustica united with Acustica}, 1955,
  \textbf{5}, 167--173\relax
\mciteBstWouldAddEndPuncttrue
\mciteSetBstMidEndSepPunct{\mcitedefaultmidpunct}
{\mcitedefaultendpunct}{\mcitedefaultseppunct}\relax
\EndOfBibitem
\bibitem[Schindelin \emph{et~al.}(2012)Schindelin, Arganda-Carreras, Frise,
  Kaynig, Longair, Pietzsch, Preibisch, Rueden, Saalfeld,
  Schmid,\emph{et~al.}]{schindelin2012fiji}
J.~Schindelin, I.~Arganda-Carreras, E.~Frise, V.~Kaynig, M.~Longair,
  T.~Pietzsch, S.~Preibisch, C.~Rueden, S.~Saalfeld, B.~Schmid \emph{et~al.},
  \emph{Nature methods}, 2012, \textbf{9}, 676--682\relax
\mciteBstWouldAddEndPuncttrue
\mciteSetBstMidEndSepPunct{\mcitedefaultmidpunct}
{\mcitedefaultendpunct}{\mcitedefaultseppunct}\relax
\EndOfBibitem
\bibitem[Bruus(2008)]{bruus2008theoretical}
H.~Bruus, \emph{Theoretical microfluidics}, Oxford university press Oxford,
  2008, vol.~18\relax
\mciteBstWouldAddEndPuncttrue
\mciteSetBstMidEndSepPunct{\mcitedefaultmidpunct}
{\mcitedefaultendpunct}{\mcitedefaultseppunct}\relax
\EndOfBibitem
\bibitem[Tinevez \emph{et~al.}(2017)Tinevez, Perry, Schindelin, Hoopes,
  Reynolds, Laplantine, Bednarek, Shorte, and Eliceiri]{tinevez2017trackmate}
J.-Y. Tinevez, N.~Perry, J.~Schindelin, G.~M. Hoopes, G.~D. Reynolds,
  E.~Laplantine, S.~Y. Bednarek, S.~L. Shorte and K.~W. Eliceiri,
  \emph{Methods}, 2017, \textbf{115}, 80--90\relax
\mciteBstWouldAddEndPuncttrue
\mciteSetBstMidEndSepPunct{\mcitedefaultmidpunct}
{\mcitedefaultendpunct}{\mcitedefaultseppunct}\relax
\EndOfBibitem
\bibitem[Mat(2019)]{MATLAB:2019b}
The Mathworks, Inc., Natick, Massachusetts, \emph{{MATLAB version R2019b}},
  2019\relax
\mciteBstWouldAddEndPuncttrue
\mciteSetBstMidEndSepPunct{\mcitedefaultmidpunct}
{\mcitedefaultendpunct}{\mcitedefaultseppunct}\relax
\EndOfBibitem
\bibitem[{COMSOL AB}()]{comsol}
{COMSOL AB}, \emph{COMSOL Multiphysics v. 5.6}, \url{www.comsol.com}\relax
\mciteBstWouldAddEndPuncttrue
\mciteSetBstMidEndSepPunct{\mcitedefaultmidpunct}
{\mcitedefaultendpunct}{\mcitedefaultseppunct}\relax
\EndOfBibitem
\bibitem[Selfridge(1985)]{selfridge1985approximate}
A.~R. Selfridge, \emph{IEEE transactions on sonics and ultrasonics}, 1985,
  \textbf{32}, 381--394\relax
\mciteBstWouldAddEndPuncttrue
\mciteSetBstMidEndSepPunct{\mcitedefaultmidpunct}
{\mcitedefaultendpunct}{\mcitedefaultseppunct}\relax
\EndOfBibitem
\bibitem[Muller \emph{et~al.}(2012)Muller, Barnkob, Jensen, and
  Bruus]{muller2012numerical}
P.~B. Muller, R.~Barnkob, M.~J.~H. Jensen and H.~Bruus, \emph{Lab on a Chip},
  2012, \textbf{12}, 4617--4627\relax
\mciteBstWouldAddEndPuncttrue
\mciteSetBstMidEndSepPunct{\mcitedefaultmidpunct}
{\mcitedefaultendpunct}{\mcitedefaultseppunct}\relax
\EndOfBibitem
\bibitem[Orosco and Friend(2021)]{orosco2021unraveling}
J.~Orosco and J.~Friend, \emph{arXiv preprint arXiv:2107.00172}, 2021\relax
\mciteBstWouldAddEndPuncttrue
\mciteSetBstMidEndSepPunct{\mcitedefaultmidpunct}
{\mcitedefaultendpunct}{\mcitedefaultseppunct}\relax
\EndOfBibitem
\bibitem[Singh \emph{et~al.}(2021)Singh, Zhang, and
  Friend]{singh2021investigation}
A.~Singh, N.~Zhang and J.~Friend, \emph{The Journal of the Acoustical Society
  of America}, 2021, \textbf{150}, 878--890\relax
\mciteBstWouldAddEndPuncttrue
\mciteSetBstMidEndSepPunct{\mcitedefaultmidpunct}
{\mcitedefaultendpunct}{\mcitedefaultseppunct}\relax
\EndOfBibitem
\bibitem[Lenshof \emph{et~al.}(2014)Lenshof, Evander, Laurell, and
  Nilsson]{lenshof2014building}
A.~Lenshof, M.~Evander, T.~Laurell and J.~Nilsson, \emph{Microscale
  Acoustofluidics}, Royal Society of Chemistry, 2014, pp. 100--126\relax
\mciteBstWouldAddEndPuncttrue
\mciteSetBstMidEndSepPunct{\mcitedefaultmidpunct}
{\mcitedefaultendpunct}{\mcitedefaultseppunct}\relax
\EndOfBibitem
\bibitem[Bruus(2012)]{bruus2012arf}
H.~Bruus, \emph{Lab on a Chip}, 2012, \textbf{12}, 1014--1021\relax
\mciteBstWouldAddEndPuncttrue
\mciteSetBstMidEndSepPunct{\mcitedefaultmidpunct}
{\mcitedefaultendpunct}{\mcitedefaultseppunct}\relax
\EndOfBibitem
\bibitem[Guti{\'e}rrez-Ramos \emph{et~al.}(2018)Guti{\'e}rrez-Ramos, Hoyos, and
  Ruiz-Su{\'a}rez]{gutierrez2018induced}
S.~Guti{\'e}rrez-Ramos, M.~Hoyos and J.~Ruiz-Su{\'a}rez, \emph{Scientific
  Reports}, 2018, \textbf{8}, 1--8\relax
\mciteBstWouldAddEndPuncttrue
\mciteSetBstMidEndSepPunct{\mcitedefaultmidpunct}
{\mcitedefaultendpunct}{\mcitedefaultseppunct}\relax
\EndOfBibitem
\bibitem[Takatori \emph{et~al.}(2016)Takatori, De~Dier, Vermant, and
  Brady]{takatori2016acoustic}
S.~C. Takatori, R.~De~Dier, J.~Vermant and J.~F. Brady, \emph{Nature
  Communications}, 2016, \textbf{7}, 1--7\relax
\mciteBstWouldAddEndPuncttrue
\mciteSetBstMidEndSepPunct{\mcitedefaultmidpunct}
{\mcitedefaultendpunct}{\mcitedefaultseppunct}\relax
\EndOfBibitem
\bibitem[Reichert \emph{et~al.}(2018)Reichert, Deshmukh, Lebovitz, and
  Dual]{reichert2018thin}
P.~Reichert, D.~Deshmukh, L.~Lebovitz and J.~Dual, \emph{Lab on a Chip}, 2018,
  \textbf{18}, 3655--3667\relax
\mciteBstWouldAddEndPuncttrue
\mciteSetBstMidEndSepPunct{\mcitedefaultmidpunct}
{\mcitedefaultendpunct}{\mcitedefaultseppunct}\relax
\EndOfBibitem
\end{mcitethebibliography}
\bibliographystyle{rsc} %the RSC's .bst file

\end{document}

\maketitle

\section{Experimental videos}

\underline{Video \texttt{video\_1\_fig\_5.mp4}:} shows the motion of $\SI{5.19}{\micro\meter}$ polystyrene particles inside the device from Fig. 5 of the main manuscript, when it is being excited with $V_{\mathrm{pp}} = \SI{22}{\volt}$ at $f = \SI{802}{\kilo\hertz}$. The video is recorded at $60 \, \mathrm{fps}$ and displayed at $30 \, \mathrm{fps}$. No external flow is applied.

% HugePiezo\_78260Hz\_50V\_CircStr
\underline{Video \texttt{video\_2.mp4}:} shows the pumping flow and the mixing flow when a device with $16.75 \times 8.74 \times \SI{17.4}{\milli\meter}$ PZT is excited at its resonance frequency - $\SI{78.26}{\kilo\hertz}$, with $V_{\mathrm{pp}} = \SI{50}{\volt}$. This demonstrates the independence of the observed pumping phenomena from the structural resonances of the silicon-glass chip and the fluidic channel resonances. The video is recorded at $60 \, \mathrm{fps}$ and displayed at $30 \, \mathrm{fps}$. No external flow is applied.

% 20201217\_002
\underline{Video \texttt{video\_3.mp4}:} shows how applying the pressure ($\SI{1.2}{\bar}$) to initiate an external flow through the main channel in combination with an excited PZT stalls the flow in any of the side channels (visualized by the suspended yeast cells). This could be exploited for medium exchange or for mixing cells from different solutions. The frequency is $\SI{790}{\kilo\hertz}$ and the applied voltage between $V_{\mathrm{pp}} = \SI{10.4}{\volt}$ and $V_{\mathrm{pp}} = \SI{7.4}{\volt}$. The video is recorded at $60 \, \mathrm{fps}$ and displayed at $30 \, \mathrm{fps}$.

\underline{Video \texttt{video\_4\_fig\_7a.mp4}:} shows the motion of $\SI{5.19}{\micro\meter}$ polystyrene particles inside the device from Fig. 7(a) of the main manuscript, when it is being excited with $V_{\mathrm{pp}} = \SI{30}{\volt}$ at $f = \SI{1.371}{\mega\hertz}$. The video is recorded and displayed at $30 \, \mathrm{fps}$. No external flow is applied.

\underline{Video \texttt{video\_5\_fig\_7b.mp4}:} shows the motion of $\SI{5.19}{\micro\meter}$ polystyrene particles inside the device from Fig. 7(b) of the main manuscript, when it is being excited with $V_{\mathrm{pp}} = \SI{6}{\volt}$ at $f = \SI{780}{\kilo\hertz}$. The video is recorded and displayed at $30 \, \mathrm{fps}$. No external flow is applied.

% 20201211\_015
\underline{Video \texttt{video\_6.mp4}:} shows the behaviour inside the symmetric chip with two side channel loops from Fig. 7(a) of the main manuscript, when it contains water with yeast cells. The focusing and the pumping phenomena closely resemble those observed with polystyrene particles in the main manuscript. The frequency is $\SI{1.438}{\mega\hertz}$ with the applied voltage of $V_{\mathrm{pp}} = \SI{30}{\volt}$. The video is recorded at $60 \, \mathrm{fps}$ and displayed at $30 \, \mathrm{fps}$. No external flow is applied.

% C3-4\_1463kHz\_30V\_30fps\_ChannelModes
\underline{Video \texttt{video\_7.mp4}:} shows the behaviour inside the asymmetric chip with two side channel loops from Fig. 7(b) of the main manuscript, when it contains water with yeast. Compared to the same device in the main manuscript (Fig. 7b), the device now features the focusing of yeast cells, due to changing the excitation frequency from $f = \SI{780}{\kilo\hertz}$ in \texttt{video\_5\_fig\_7b.mp4} to $f = \SI{1.463}{\mega\hertz}$, where the resonance of the fluidic channel is excited. The voltage applied to the PZT is $V_{\mathrm{pp}} = \SI{30}{\volt}$. The video is recorded at $60 \, \mathrm{fps}$ and displayed at $30 \, \mathrm{fps}$. No external flow is applied.

% 20201214\_005
%\underline{Video \texttt{video\_8.mp4}:} shows yeast cells suspended in water in one of the devices, demonstrating the possibility to continuously control the pumping flow intensity by linearly sweeping the excitation voltage, from $V_{\mathrm{pp}} = \SI{30}{\volt}$ to $\SI{30}{\volt}$ and back to $\SI{30}{\volt}$. The frequency of excitation is $f = \SI{667}{\kilo\hertz}$. The video is recorded at $60 \, \mathrm{fps}$ and displayed at $30 \, \mathrm{fps}$.

\underline{Video \texttt{video\_8\_fig\_8.mp4}:} shows the trapping of yeast cells suspended in water inside the device from Fig. 8 of the main manuscript, when it is being excited with $V_{\mathrm{rms}} = \SI{30}{\volt}$ at $f = \SI{714}{\kilo\hertz}$. The video is recorded at $60 \, \mathrm{fps}$ and displayed at $30 \, \mathrm{fps}$. No external flow is applied.

%Picture of the sharp edge tip rounding.

\section{Perturbed equations}

\subsection*{First-order (acoustic) problem}
For a quiescent fluid at the zeroth order ($\boldsymbol{v}_0 = \boldsymbol{0}$), the substitution of the perturbed fields into the governing equations yields the following set of first-order equations,
\begin{align}
\rho_0 \frac{\partial \boldsymbol{v}_1}{\partial t} &= - \boldsymbol{\nabla} p_1 + \eta \nabla^2 \boldsymbol{v}_1 + \left( \eta_{\mathrm{B}} + \frac{\eta}{3} \right) \boldsymbol{\nabla} \left( \boldsymbol{\nabla} \boldsymbol{\cdot} \boldsymbol{v}_1 \right) \label{al:eq1} ,\\
\frac{\partial \rho_1}{\partial t} &= -\rho_0 \boldsymbol{\nabla} \boldsymbol{\cdot} \boldsymbol{v}_1 \label{al:eq2} ,
\end{align}
with the equilibrium density $\rho_0$. The equation of state for the barotropic fluid,
\begin{equation}
\rho_1 = \frac{1}{c_0^2} p_1,
\label{eq:state}
\end{equation}
is connecting the first-order density with the first-order pressure through the speed of sound in the fluid $c_0$. The first-order fields are assumed to have a harmonic time-dependency with $e^{\mathrm{i} \omega t}$, where $\omega = 2 \pi f$, with the frequency $f$.

\subsection*{Second-order (streaming) problem}
Applying the perturbation theory up to second order to the governing equations, namely
\begin{align}
    \rho_0 \frac{\partial \boldsymbol{v}_2}{\partial t} &- \left( \eta_{\mathrm{B}} + \frac{\eta}{3} \right) \boldsymbol{\nabla} \left( \boldsymbol{\nabla} \boldsymbol{v}_2 \right) + \boldsymbol{\nabla} p_2 - \eta \nabla^2 \boldsymbol{v}_2 = \nonumber\\
    & \quad - \rho_0 ( \boldsymbol{v}_1 \boldsymbol{\cdot} \boldsymbol{\nabla} ) \boldsymbol{v}_1 -\rho_1 \frac{\partial \boldsymbol{v}_1}{\partial t} \label{td:NS}
\end{align}
and
\begin{equation}
    \frac{\partial \rho_2}{\partial t} + \rho_0 \boldsymbol{\nabla} \boldsymbol{\cdot} \boldsymbol{v}_2 = -\rho_1 \boldsymbol{\nabla} \boldsymbol{\cdot} \boldsymbol{v}_1 \label{td:cont} ,
\end{equation}
together with taking the time average $\langle \square \rangle = \frac{1}{T}\int_T \square \mathrm{d}t $ over an oscillation period $T$, results in the equations of acoustic streaming,\cite{doinikov1994acoustic,lighthill1978acoustic}
\begin{align}
    \boldsymbol{\nabla} \left<p_2\right>- \eta \nabla^2 \left< \boldsymbol{v}_2 \right> &-\left( \eta_{\mathrm{B}}+\frac{\eta}{3} \right) \boldsymbol{\nabla} \left(\boldsymbol{\nabla} \boldsymbol{\cdot} \left< \boldsymbol{v}_2 \right> \right) \nonumber \\
    &\qquad \qquad= - \rho_0 \boldsymbol{\nabla} \boldsymbol{\cdot} \left< \boldsymbol{v}_1 \boldsymbol{v}_1 \right>, \label{al:PartStramIII} \\
    \rho_0 \boldsymbol{\nabla} \boldsymbol{\cdot} \left< \boldsymbol{v}_2 \right> &= -\boldsymbol{\nabla} \boldsymbol{\cdot} \left<\rho_1 \boldsymbol{v}_1\right>. \label{al:PartStramIV}
\end{align}
At the second order, the no-slip boundary condition is imposed on the Lagrangian fluid velocity at the fluid-solid interface, to account for the oscillations of the interface at the first order. The Lagrangian velocity is defined as the summation of the Eulerian streaming velocity $\left< \boldsymbol{v}_2 \right>$ and the Stokes drift \cite{andrews1978exact, buhler2014waves}
\begin{equation}
    \boldsymbol{v}_{\mathrm{SD}} = \left< \left( \int{\boldsymbol{v}_1}\mathrm{d}t \boldsymbol{\cdot} \boldsymbol{\nabla} \right) \boldsymbol{v}_1 \right>.
\end{equation}
The boundary condition consequently translates into
\begin{equation}
\left< \boldsymbol{v}_2 \right> = - \boldsymbol{v}_{\mathrm{SD}} \quad \text{at the interface}.
\end{equation}
However, $\boldsymbol{v}_{\mathrm{SD}}=0$ at the boundaries that are rigid.

\subsection*{Motion of an object}
In addition to the ARF, the object in an acoustic field is subjected to the Stokes drag\cite{bruus2012arf} due to the acoustic streaming velocity,
\begin{equation}
  \boldsymbol{F}_{\mathrm{s}} = 6 \pi \eta a \left( \left< \boldsymbol{v}_2 \right> - \boldsymbol{v}_{\mathrm{obj}} \right),
  \label{eq:Fstream}
\end{equation}
where $\boldsymbol{v}_{\mathrm{obj}}$ is the velocity of the object.

Combining both force contributions, namely the Stokes drag and the ARF, the overall motion of an object with mass $m$ is governed by the Newton's second law
\begin{equation}
  m \frac{\mathrm{d} \boldsymbol{v}_{\mathrm{obj}}}{\mathrm{d} t} = \boldsymbol{F}_{\mathrm{s}} + \boldsymbol{F}_\mathrm{rad}.
  \label{eq:motion}
\end{equation}
For length scales applicable in our work, it is sufficient to approximate the motion of such an object by assuming it reaches its terminal velocity instantaneously,\cite{bach2020suppression} resulting in the force equilibrium
\begin{equation}
  \boldsymbol{F}_\mathrm{rad} = - \boldsymbol{F}_{\mathrm{s}}.
  \label{eq:equilibriumForce}
\end{equation}
The velocity of the object, subjected to the drag force and the ARF, can then be estimated as
\begin{equation}
  \boldsymbol{v}_{\mathrm{obj}} = \frac{\boldsymbol{F}_\mathrm{rad}}{6 \pi \eta a} + \left< \boldsymbol{v}_2 \right> .
  \label{eq:velocityObj}
\end{equation}

Assuming a one-dimensional standing wave and substituting eq. (5) from the main manuscript as an approximation for the ARF into eq. (\ref{eq:velocityObj}), neglecting the streaming velocity, and performing integration, results in the trajectory\cite{bruus2012arf} of the object,
\begin{equation}
  z(t) = \frac{1}{k} \arctan \left\{ \tan \left[ k z(0) \right] \exp{\left[\frac{4 \Phi E_{\mathrm{ac}}}{3 \eta} ( k a )^2 t \right] } \right\}.
  \label{eq:traject}
\end{equation}

\section{Numerical model}

The first-order acoustic problem, as defined with equations (\ref{al:eq1}), (\ref{al:eq2}), and (\ref{eq:state}), was solved via the adiabatic formulation of the Thermoviscous Acoustics interface, combined with the Frequency Domain study. The resulting acoustic fields were then applied as a source term to the second-order streaming problem that is defined with equations (\ref{al:PartStramIII}) and (\ref{al:PartStramIV}). The streaming problem was implemented by adjusting the Creeping Flow interface to include the source terms from the first-order study as a volume force. Stationary study was used to solve the equations of the streaming problem, which is a valid approach under the assumption that the flow is time-independent. 

\begin{figure}[h]
\centering
 \includegraphics{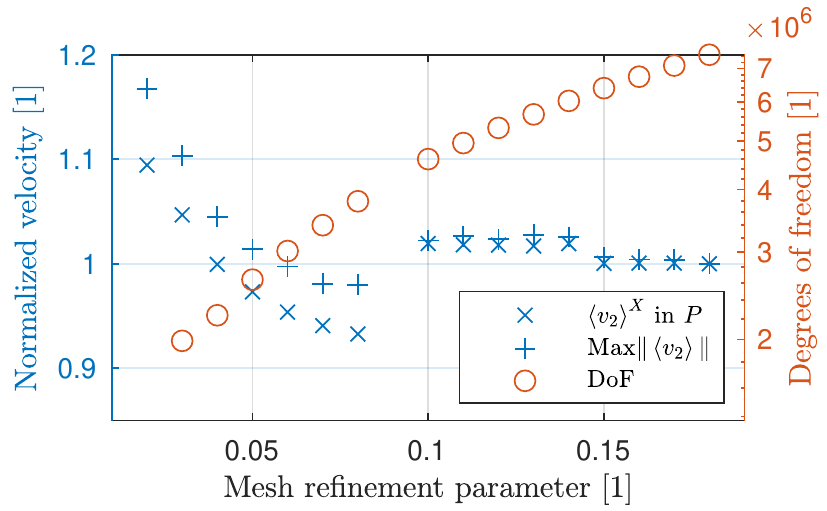}
 \caption{Mesh convergence study to determine the necessary mesh refinement in the viscous boundary layer, at $f=\SI{746}{\kilo\hertz}$. The left vertical axis shows the normalized velocity, while the right vertical axis shows the corresponding degrees of freedom.}
 \label{fig:sup2}
\end{figure}

The convergence of the numerical model with respect to a decreasing size of mesh elements is shown in Fig. \ref{fig:sup2}, for the simulation parameters used in Fig. 4 of the main manuscript ($f= \SI{746}{\kilo\hertz}$).

% \begin{figure}[h]
%  \centering
%  \includegraphics{figures/frequency_study.pdf}
%  \caption{Frequency sweep and analysis of the acoustic energy density and streaming velocities.}
%  \label{fig:freq_sweep}
% \end{figure}

\section{Admittance analysis}
\begin{figure}[h]
\centering
 \includegraphics{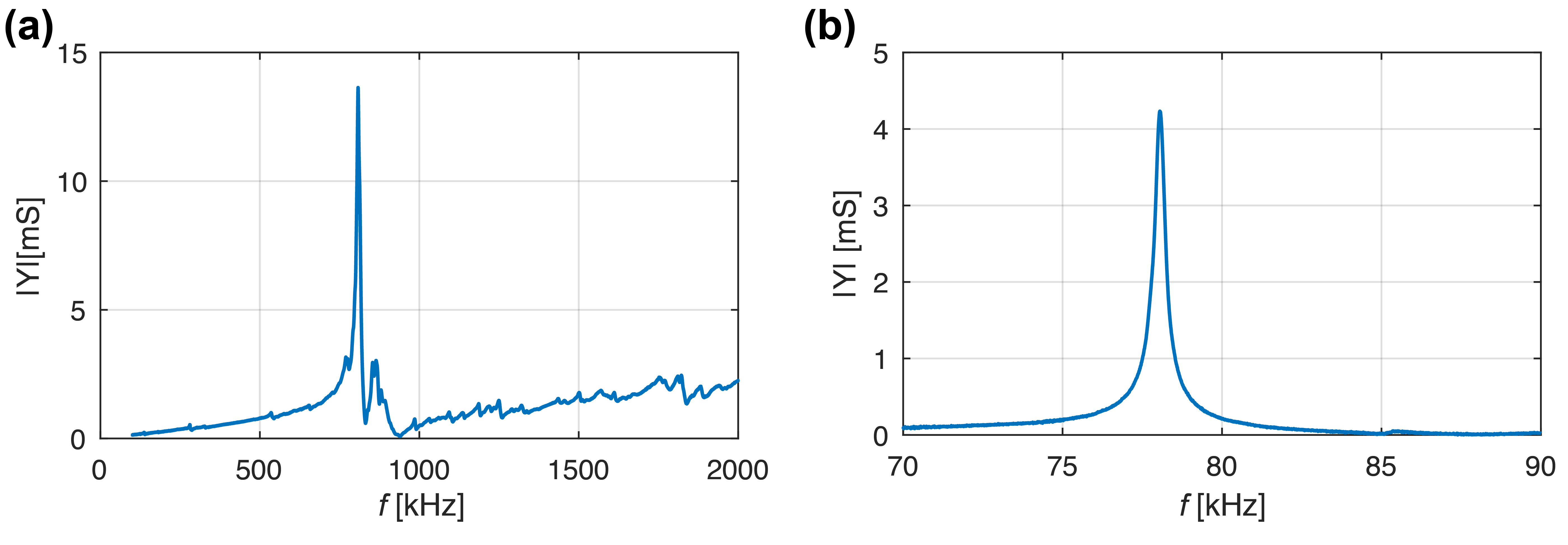}
 \caption{(a) Admittance measured with respect to the excitation frequency for the $10\times2\times\SI{1}{\milli\meter}$ PZT on the device from Fig. 4 of the main manuscript. (b) Admittance measured with respect to the excitation frequency for the $16.75\times8.74\times\SI{17.4}{\milli\meter}$ PZT on the device from \texttt{video\_2.mp4}.}
 \label{fig:sup3}
\end{figure}

Figure \ref{fig:sup3}(a) shows the admittance curve of the $10\times2\times\SI{1}{\milli\meter}$ PZT attached to the device from Fig. 4 of the main manuscript. The curve indicates a resonance of the PZT at around $\SI{800}{\kilo\hertz}$, which corresponds to the frequency where a strong pumping flow and cell/microparticle focusing was observed.

Figure \ref{fig:sup3}(b) shows the admittance curve of the $16.75\times8.74\times\SI{17.4}{\milli\meter}$ PZT attached to the device from \texttt{video\_2.mp4}. The curve indicates a resonance of the PZT at around $\SI{78}{\kilo\hertz}$, corresponding to the frequency where a strong pumping flow was observed.

% \begin{figure}[h]
% \centering
%  \includegraphics{figures/hugePiezo_admittance.pdf}
%  \caption{Admittance measured with respect to the excitation frequency for the $16.75\times8.74\times\SI{17.4}{\milli\meter}$ PZT on the device from \texttt{video\_2.mp4}.}
%  \label{fig:sup4}
% \end{figure}

\section{Determining acoustic pressure amplitude from particle trajectories}

\begin{figure}[h]
 \centering
 \includegraphics[scale=1]{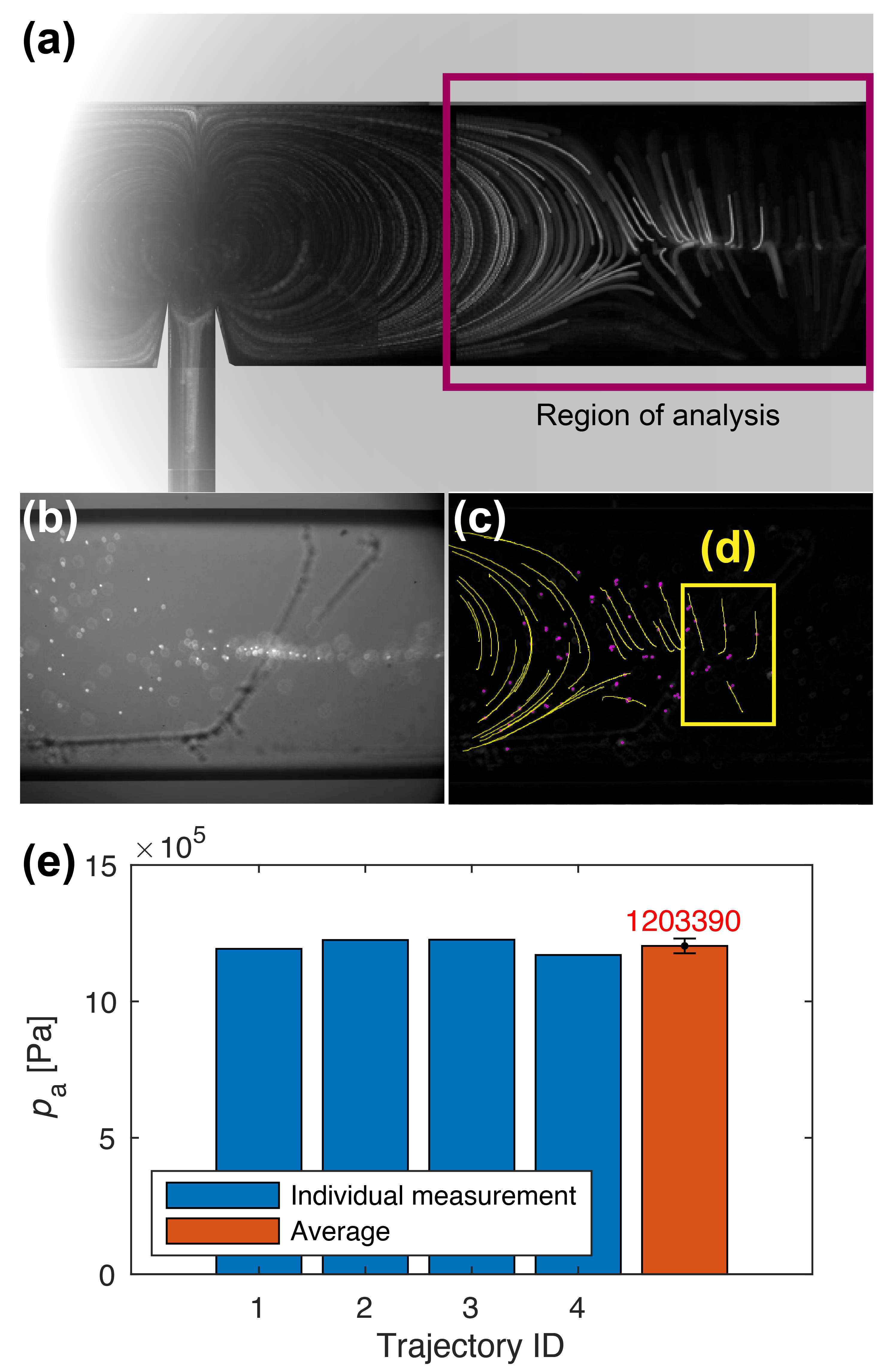}
 \caption{The acoustic pressure amplitude of a standing wave responsible for the focusing in the device from Fig. 5 of the main manuscript, assessed based on the region (a) that shows minimal influence of the pumping and the vortices. (b) The focused $\SI{5.19}{\micro\meter}$ polystyrene particles, at $f=\SI{802}{\kilo\hertz}$ and $V_{\mathrm{pp}} = \SI{22}{\volt}$. (c) The particle trajectories based on the particle tracking velocimetry approach, and (d) the trajectories that are isolated from the mixing vortex and used for the pressure amplitude analysis. (e) The extracted acoustic pressure amplitude $p_{\mathrm{a}}$ for each trajectory, and the resulting average of $\SI{1.2}{\mega\pascal}$. Magenta-colored dots in (c) are the outlines of PS particles midway through the focusing.}
 \label{fig:p_a_estimate}
\end{figure}

The acoustic pressure amplitude was measured in the region of the main channel outlined in Fig. \ref{fig:p_a_estimate}(a)-(c) that did not experience the mixing or the pumping flow, as it was outside the pumping flow loop.
The pressure amplitude was computed by fitting the PTV trajectories to the theoretical particle trajectory from eq. (\ref{eq:traject}), using $p_{\mathrm{a}}$ as the fitting parameter.\cite{barnkob2010measuring} For the calculation, we selected the four longest PTV trajectories, the results of which are displayed in Fig. \ref{fig:p_a_estimate}(e), with the average pressure amplitude of $\SI{1.2}{\mega\pascal}$.

%\section{Flow rate and pumping pressure computation}

%Describe the flow rate computation procedure, as well as the computation of the pumping pressure from the flow rate.

\clearpage
%%%REFERENCES%%%
\bibliography{rsc} %You need to replace "rsc" on this line with the name of your .bib file
\bibliographystyle{rsc} %the RSC's .bst file